\documentclass[10pt,preprint,aps]{revtex4-2}
\usepackage[latin9]{inputenc}
\usepackage[active]{srcltx}
\usepackage{xcolor}
\usepackage{amsmath}
\usepackage{graphicx}
\usepackage{rotfloat}
\PassOptionsToPackage{normalem}{ulem}
\usepackage{ulem}
\usepackage{appendix}

\makeatletter

\providecolor{lyxadded}{rgb}{0,0,1}
\providecolor{lyxdeleted}{rgb}{1,0,0}

\DeclareRobustCommand{\lyxsout}[1]{\ifx\\#1\else\sout{#1}\fi}

\usepackage{lineno}
\usepackage{hyperref}

\makeatother

\begin{document}

\title{\textcolor{black}{
Prospects for the determination of fundamental constants with beyond-state-of-the-art uncertainty using molecular hydrogen ion spectroscopy}}

\author{S.\,Schiller{*}}
\affiliation{Institut f\"ur Experimentalphysik, Heinrich-Heine-Universit\"at
D\"usseldorf, 40225 D\"usseldorf, Germany}
\author{J.-Ph.\,Karr$^{1,2}$}
\affiliation{{$^1$}{Laboratoire Kastler Brossel, Sorbonne Universit\'e, CNRS, ENS-Universit\'e PSL,
Coll\`{e}ge de France, 4 place Jussieu, F-75005 Paris, France;}
{$^2$}{Universit\'e d\textquotesingle Evry-Val d\textquotesingle Essonne, Universit\'e Paris-Saclay, Boulevard Fran\c{c}ois Mitterrand, F-91000 Evry, France}}

\begin{abstract}

\textbf{The proton, deuteron and triton masses can be determined
relative to the electron mass via rovibrational spectroscopy of molecular
hydrogen ions. This 
has to occur
via comparison of the experimentally measured
transition frequencies and the \emph{ab initio} calculated frequencies, whose dependence on the mass ratios can be
calculated precisely. In 
{precision} experiments to date 
{(on HD$^+$ and H$_2^+$)}, the transitions have
involved the ground vibrational level $v=0$ and excited vibrational
levels with quantum numbers up to $v'=9$. For these transitions, the
sensitivity of the }\textbf{\emph{ab initio}}\textbf{ frequency on
the high-order-QED contributions is correlated with that on the mass
ratios. 
This prevents an efficient simultaneous determination of these quantities from experimental data, so that the accuracy of the mass ratios is essentially limited by the theoretical uncertainty.
Here we analyze how the accuracy of mass ratios may
be improved by providing experimental transition frequencies between
levels with larger quantum numbers, whose sensitivity on the mass ratio
is positive rather than negative, or close to zero.
This allows the unknown QED contributions and involved fundamental constants to be much more efficiently determined from a joint analysis of several measurements.
We also consider scenarios where transitions of D$_2^+$ are included.
We find these to be powerful approaches, allowing 
in principle to reach uncertainties for the mass ratios approximately three orders smaller than CODATA 2018. 
Improvements by a factor of 3.5 for the Rydberg constant, and 11 (14) for the proton (deuteron) charge radius, are also projected.
}
\end{abstract}
\maketitle

\section{Introduction}

The precision spectroscopy of the molecular hydrogen ion (MHI) HD\textsuperscript{+}
has made significant progress in the past few years. The combination
of recently obtained experimental \citep{Alighanbari2020,Germann2021,Kortunov2021,Alighanbari2023}
and theoretical results \citep{Korobov2021} on the transition frequencies
permits extracting one ``fundamental'' constant, the reduced proton-deuteron
mass relative to the electron mass, $\mu_{pd}/m_{e}$. The current uncertainty from MHI spectroscopy, $u([\mu_{pd}/m_e]_{{\rm exp,HD}^+})\simeq2.5\times10^{-8}$~\citep{Alighanbari2023,Karr2023}, is competitive with direct mass measurements using Penning traps~\citep{Koehler2015,Heisse2019,Rau2020,Fink2021}. 
The accuracies of some of the experimentally
determined transition frequencies are already higher than those of
the theoretical predictions. Hence, the latter limit the accuracy
of the determination of $\mu_{pd}/m_{e}$. The theoretical uncertainty
is dominated by unknown high-order-QED contributions \citep{Korobov2021}.

The question arises which are the fundamental limitations
to the attainable uncertainties in the determination of the mass ratios, and of
other constants,  using {\em any possible future  result} of MHI spectroscopy. The question has first
been addressed in an earlier analysis \citep{Karr2016a}, where different
measurement scenarios were considered. A scenario that included three
HD\textsuperscript{+} transitions and two ${\rm H}_{2}^{+}$ transitions (assumed to have been measured with fractional uncertainties $1\times10^{-12}$), no other experimental input data, and assumed $3\times10^{-12}$ theory uncertainty,
resulted in a proton-electron mass ratio uncertainty $u(m_{p}/m_{e})\simeq1.5\times10^{-8}$.

In the present study, we seek to ``break'' the 
correlation between the theoretical uncertainties of the transition frequencies due to uncalculated terms. The existence of
this correlation has first been emphasized in ref.~\citep{Karr2016a}.
Alighanbari et al.\,\citep{Alighanbari2023} observed that the correlation is partially removed when computing ratios of theoretical transition frequencies between levels of not-too-disparate vibrational quantum numbers. The agreement between theoretical and experimental ratios exhibits a particularly small combined uncertainty.


Two different approaches are used in our study: (i) a simplified approach, in which the unknown QED contributions are treated as a single parameter to be determined by solving a system of linear equations. (ii) a least-squares adjustment (LSA), similar to CODATA~\citep{Mohr2000} and Ref.~\citep{Karr2023}, where QED contributions to the different transition frequencies are treated as adjustable parameters. 
We also propose to include qualitatively different vibrational
transitions in the measurement program 
and discuss several scenarios, i.e.\,sets of transitions measured on different MHI.

\begin{figure}[h!]
\begin{centering}
\includegraphics[width=.7\columnwidth]{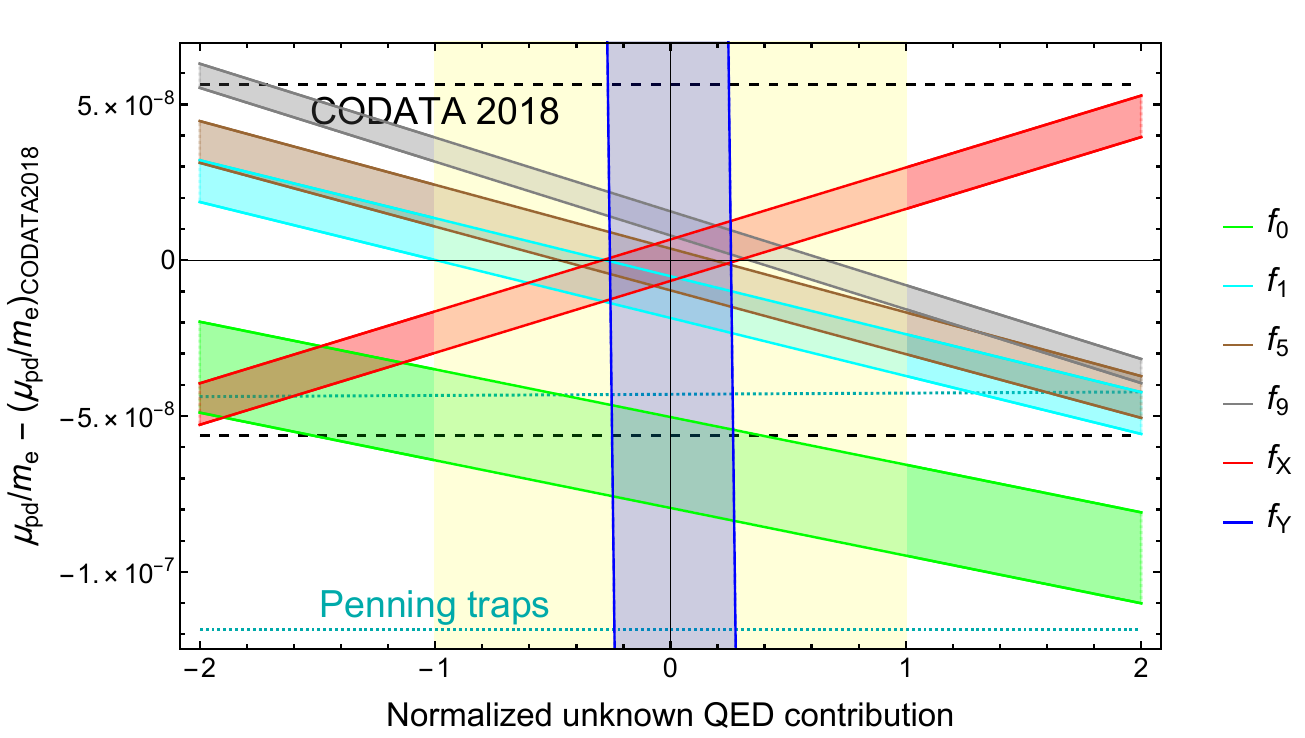}
\par
\end{centering}
\caption{\label{fig:Four-HD-transitions}
{\footnotesize Determination of the nuclear-electron mass
ratio of HD\protect\textsuperscript{+} from single transition frequencies.
Shown are four already measured transitions ($f_{0},$ $f_{1},$ $f_{5}$,
$f_{9}$) and two \textquotedblleft hot-band\textquotedblright{} transitions
considered for future measurement, $X:(v=9,N=1)\rightarrow$ $(v'=18,N'=0)$,
$Y:(7,1)\rightarrow(15,0)$. $f_{X}$ (red) is a \textquotedblleft positive-mass-sensitivity\textquotedblright{}
transition, $f_{Y}$ (blue) is a \textquotedblleft suppressed-mass-sensitivity\textquotedblright{}
transition. Note that the band slope is similar for $f_{0},$ $f_{1},$ $f_{5}$,
$f_{9}$ because the abscissa is the normalized QED contribution $\Delta_{\rm nuc,QED}/0.32$, see eq.~(\ref{eq:master equation}). The width of each band takes into account (1) the uncertainty
of the theoretical transition frequency due to the uncertainty of
the Rydberg constant and the uncertainties of the r.m.s. charge radius
of proton and deuteron, (2) the uncertainty of the experimentally
measured frequency, (3) the uncertainty of the hyperfine structure
theory. 
For the four measured transitions, the experimental results
$f^{\rm (exp)}$ and theoretical results with CODATA 2018 constants $f_{{\rm 2018}}^{\rm (theor)}$ have been used in
the plot. For the proposed transitions (red and blue bands) we have
assumed a hypothetical perfect match, $f^{\rm (exp)}=f_{{\rm 2018}}^{\rm (theor)}$,
experimental uncertainties $u(f_{X}^{\rm (exp)})=u(f_{Y}^{\rm (exp)})=0.3\,{\rm kHz},$
and 0.035~kHz spin theory uncertainty. 
The yellow shaded range indicates
the estimated normalized uncertainty of the unknown QED contribution
to the theoretical transition frequency, stemming from the three contributions
$u(\delta f_{{\rm QED},1})$, $u(\delta f_{{\rm QED},2})$, $u(\delta f_{{\rm QED},3})$
(see text for details). 
The horizontal black dashed lines indicate the $\pm1$ standard uncertainty interval
of the value of $\mu_{pd}/m_{e}$ according to CODATA~2018. The dark cyan
dotted lines represent the $\pm1$ standard uncertainty interval of $\mu_{pd}/m_{e}$
derived from Penning trap measurements. It is computed from the uncertainties
of the atomic mass of the electron \citep{Koehler2015} (modified
as in \citep{Tiesinga2021}), of the atomic mass of the deuteron \citep{Rau2020},
and of the proton-deuteron mass ratio \citep{Fink2021}. 
}}
\end{figure}

Figure~\ref{fig:Four-HD-transitions} summarizes the current situation
and presents one main idea of this paper. The figure displays, for
six transitions of HD\textsuperscript{+}, the correlation between
the unknown (uncalculated) QED contribution and the mass ratio value
deduced from requiring the theoretical frequency to match the experimental
frequency. The four transition frequencies measured to date (green,
light blue, brown and gray) have a similar relationship between mass
uncertainty and QED uncertainty: the bands have similar slopes. Considering
each measured frequency independently, as well as the estimated uncertainty
of the QED contributions (yellow band), one can derive a value for
the mass ratio with uncertainty $
u([\mu_{pd}/m_e]_{{\rm exp,HD}^+})\simeq2.5\times10^{-8}$. 
This is substantially determined by the QED uncertainty.

The similar band slopes imply that even when the four transitions
are taken together, they are not particularly effective in reducing
the uncertainty of the mass ratio. If, however, a suitably chosen
additional transition frequency is introduced, whose slope is qualitatively
different ($f_{X}$, red band), then a simultaneous determination
of both the mass correction and the QED correction with reduced uncertainty
becomes possible. This will be discussed in detail in the following.

\eject
\newpage

\subsection{Theoretical transition frequencies}

We only consider the so-called spin-averaged rovibrational frequencies.
This is reasonable since it has been shown theoretically that the
contribution of hyperfine energies can be completely eliminated by
measuring the complete set of hyperfine components of a transition
and then applying a sum rule \citep{Schiller2018}. A first demonstration
has recently been given \citep{Schenkel2023}. The {\em ab initio} theory
of the spin-averaged frequency is well-developed \citep{Korobov2021}.
A spin-averaged theoretical frequency may be decomposed as

\begin{eqnarray}
    f^{\rm (theor)} & =&f_{nr}(R_{\infty},\{m_{i}/m_{e}\})+\delta f_{\rm QED}(R_{\infty},\alpha,\{m_{i}/m_{e}\}) +\\ \nonumber
 & \phantom{=}&+\delta f_{\rm nuc} (\{r_{i}\})+\delta f_{\rm nuc,h.o.}\,,
\label{eq:f_theor}
\end{eqnarray}
where the important dependencies on fundamental constants are indicated.
Here, $R_{\infty}$ is the Rydberg constant, $m_{i}$ is the mass of a nucleus
 (proton, deuteron, triton), $m_{e}$ is the electron mass, $\alpha$
is the fine-structure constant, $r_{i}$ is the charge radius
of proton, deuteron, or triton. In addition, all terms in the equation
are functions of the vibrational and rotational quantum numbers of
the lower $(v,N)$ and upper $(v',N')$ rovibrational levels; they
will be indicated explicitly below. The first term, $f_{nr}$, is the
nonrelativistic transition frequency, arising from the solution of
the three-body Schr\"odinger equation. The second term, $\delta f_{\rm QED}$, contains all energy corrections due to relativistic and QED effects. The third term is the leading-order finite-nuclear-size shift. Finally, the fourth term, $\delta f_{\rm nuc,h.o.}$, contains small nuclear
corrections of higher order, whose dependencies on fundamental constants may be neglected. The fundamental constants displayed in the equation above represent the dominant dependencies.

MHI spectroscopy can be exploited to determine one or more of the above fundamental
constants from the comparison of a set of experimentally measured
transition frequencies $\{f_{k}^{\rm (exp)}\}$ and their theoretical
counterparts $\{f_{k}^{\rm (theor)}\}$. We therefore first discuss qualitatively
the impact of theoretical uncertainties and uncertainties of the fundamental
constants on the four terms of $f_{k}^{\rm (theor)}$.

\subsection{Uncertainty of the theoretical prediction} \label{sec:theory uncertainty}

To begin with, we consider the uncertainties stemming from the involved fundamental constants. All terms in eq.\,(\ref{eq:f_theor}) are proportional to the Rydberg constant; however the impact
of its current uncertainty (CODATA 2018 \cite{Tiesinga2021}) is mainly on the first term,
since the second-largest term, $\delta f_{\rm QED}$, is approximately
$5\times10^{4}$ times smaller than the first. The uncertainty of the mass ratios therefore
also affects mostly the first term. Since we actually wish to determine
mass ratios more accurately than provided by CODATA 2018, the sensitivity
of $f_{nr}$ to the mass ratios, i.e.\,the partial derivatives $\partial f_{nr}/\partial(m_{i}/m_{e})$,
are important quantities. They have been computed in a number of works,
e.g. \citep{Schiller2005,Karr2006}. Since the fine-structure constant
does not enter $f_{nr}$ except via $R_\infty$, its current uncertainty is not relevant in $f_{nr}$.

The third term $\delta f_{\rm nuc}$ is equal to the difference of the finite-nuclear-size shifts in the upper and lower levels. The shift of each energy level is proportional to the expectation
value of the delta potential $V_{\delta,i}=\delta({\bf r}_{e,i})$
of the electron at nucleus $i$, i.e. to the probability density at
the nucleus. $\langle V_{\delta,i}\rangle$ is expressed in atomic
units
. Its values can be accurately calculated.
The nuclear-size shift for a heteronuclear MHI is given by \citep{Aznabayev2019}
\begin{equation}
\delta f_{\rm nuc}=2cR_{\infty}(2\pi/3)a_{0}^{-2}\left(r_{1}^{2}(\langle V_{\delta,1}\rangle_{v',N'}-\langle V_{\delta,1}\rangle_{v,N})+r_{2}^{2}(\langle V_{\delta,2}\rangle_{v',N'}-\langle V_{\delta,2}\rangle_{v,N})\right)\,,\label{eq:f_nuc}
\end{equation}
where $r_{1}$, $r_{2}$ are the 
charge
radii of the two nuclei. The value of $\delta f_{\rm nuc}$ is uncertain
because of the uncertainties of $r_{1},r_{2}$. The density values
$\langle V_{\delta,i}\rangle_{v,N}$ decrease as the vibration becomes
more excited (larger $v$), because the molecule becomes more stretched.
For HD$^{+}$, $\langle V_{\delta,i}\rangle_{v,N=0}$ varies from $\approx0.21$
for $v=0$ to $\approx0.17$ for $v=9$ to $\approx0.16$ for $v=18$.
For the following, it is important to note that the values for proton
and deuteron  are close, differing only by $\le0.14$\% for $v\le9$ and by 1.5\% for
$v=18$.

The following nuclear charge radius data have been recommended by CODATA
2018~\citep{Tiesinga2021}:
\begin{align}
r_{\text{p,2018}}^{2} & =0.7080(32)\,\text{fm}^{2},\nonumber \\
r_{\text{d,2018}}^{2} & =4.5283(31)\,\text{fm}^{2}\,.\label{eq:rp2 and rd2 from CODATA18}
\end{align}
The two uncertainties are almost perfectly correlated, because the deuteron-proton radius difference $r_d^2-r_p^2$ is strongly constrained by the measurement of the isotope shift of 1S-2S transition in H and D. Its recommended value is
\begin{equation}
[r_{d}^{2}-r_{p}^{2}]_{2018}=3.82036(41)\,\text{fm}^{2}.\label{eq:rd2-rp2}
\end{equation}
We can reexpress eq.~(\ref{eq:f_nuc}) as
\begin{eqnarray}
\delta f_{\rm nuc}
&=&2cR_{\infty}(2\pi/3)a_{0}^{-2}\frac{1}{2}\Bigl(\left(r_{1}^{2}+r_{2}^{2}\right)(\langle V_{\delta,1}\rangle_{v',N'}+\langle V_{\delta,2}\rangle_{v',N'}-(\langle V_{\delta,1}\rangle_{v,N}+\langle V_{\delta,2}\rangle_{v,N}))+\nonumber
\\
 & &\phantom{2cR_{\infty}(2\pi/3)a_{0}^{-2}\frac{1}{2}}
 \left(r_{1}^{2}-r_{2}^{2}\right)\left(\langle V_{\delta,1}\rangle_{v',N'}-\langle V_{\delta,2}\rangle_{v',N'}-(\langle V_{\delta,1}\rangle_{v,N}-\langle V_{\delta,2}\rangle_{v,N})\right)\biggr)\nonumber
 \\
 &=&2cR_{\infty} \frac{2\pi}{3}a_{0}^{-2}\frac{1}{2}
 \Bigl(
 \left(r_{1}^{2}\!+\!r_{2}^{2}\right)\left(\langle V_{\delta,12}\rangle_{v',N'}\!-\!\langle V_{\delta,12}\rangle_{v,N}\right)
 +\nonumber\\
 & & \phantom{2cR_{\infty} \frac{2\pi}{3}a_{0}^{-2}\frac{1}{2} \Bigl(}
 \left(r_{1}^{2}\!-\!r_{2}^{2}\right)
 \left(\langle \Delta V_{\delta,12}\rangle_{v',N'}\!-\!\langle \Delta V_{\delta,12}\rangle_{v,N}\right)
 \Bigr)\,.
 \label{delta_f_nuc}
\end{eqnarray}
with the notations $\langle V_{\delta,12}\rangle=\langle V_{\delta,1}+V_{\delta,2}\rangle$
and $\langle \Delta V_{\delta,12}\rangle=\langle V_{\delta,1}-V_{\delta,2}\rangle$.

For homonuclear MHI the second term
in the parenthesis is zero by definition. In order to keep the analytical model simple, for heteronuclear MHI we shall ignore the nominal value of the second term, 
i.e. its value for $r_i=r_{i,2018}$. For actual computations it could easily be included. We shall also neglect its uncertainty, which can be justified as follows. 
In a given fit scenario, one obtains the order of magnitude of the uncertainty of the combination $r_1^2+r_2^2$ appearing in the first term. The uncertainty of the combination $r_1^2-r_2^2$ will  be similar or smaller for some scenarios (e.g. when both HD$^+$ and H$_2^+$ data is included). In any case, the multiplying factor $\langle \Delta V_{\delta}\rangle_{v',N'}-\langle \Delta V_{\delta}\rangle_{v,N}$ is a factor of at least 6 smaller than $\langle  V_{\delta}\rangle_{v',N'}-\langle V_{\delta}\rangle_{v,N}$, see column~7 in Tab.\,\ref{tab:Properties of transitions}. Thus, we may neglect the uncertainty of the second term in the parenthesis compared to the uncertainty of the first term.
One may improve on this approximation in an extended model, where one incorporates the experimental results from H/D spectroscopy, i.e. the result eq.~(\ref{eq:rd2-rp2}) together with its uncertainty, in the second term. Its uncertainty will then be negligible. 

Thus, the uncertainty of $\delta f_{\rm nuc}$
is approximately 
\begin{align}
u(\delta f_{\rm nuc}) & \simeq2cR_{\infty}(2\pi/3)a_{0}^{-2}\frac{1}{2}
u(r_{1,2018}^{2}+r_{2,2018}^{2})
\left(\langle V_{\delta,12}\rangle_{v',N'}-\langle V_{\delta,12}\rangle_{v,N}\right)\nonumber \\
 & =15.6\,{\rm kHz}\times(\langle V_{\delta,pd}\rangle_{v',N'}-\langle V_{\delta,pd}\rangle_{v,N})\,\,\,\,\,\,\,[{\rm HD}^{+}] \\
  & =15.7\,{\rm kHz}\times(\langle V_{\delta,pp}\rangle_{v',N'}-\langle V_{\delta,pp}\rangle_{v,N})\,\,\,\,\,\,\,[{\rm H}_{2}^{+}] \, .
\label{eq: u(delta f_nuc)}
\end{align}
In the second line, for HD$^+$, we inserted the uncertainty 
$u(r_{p,2018}^{2}+r_{d,2018}^{2})=$ $0.0063\,{\rm fm}^{2}$,
where correlation has been taken into account.

We now discuss uncertainties of purely theoretical origin. They mainly concern the QED correction $\delta f_{\rm QED}$ and arise dominantly
from QED contributions of high order in $\alpha$ ($R_{\infty} \alpha^6$ and above). An important point is that the largest sources of uncertainties are described by terms that
are also proportional to $\langle V_{\delta,12}\rangle_{v',N'}-\langle V_{\delta,12}\rangle_{v,N}$
\citep{Korobov2021}. The largest one is the higher-order remainder
of the one-loop self-energy correction, which has an estimated uncertainty
\begin{align}
u(\delta f_{{\rm QED},1}) & =2cR_{\infty}\times41.2\,\alpha^{6}(\langle V_{\delta,12}\rangle_{v',N'}-\langle V_{\delta,12}\rangle_{v,N})\,\nonumber \\
 & =40.9\,{\rm kHz}\times(\langle V_{\delta,12}\rangle_{v',N'}-\langle V_{\delta,12}\rangle_{v,N})\,.\label{eq:u(delta_f_QED1)}
\end{align}
The second largest source of uncertainty is the higher-order remainder of the two-loop QED correction. The uncertainty associated with this term is 
\begin{align}
u(\delta f_{{\rm QED},2}) & =2cR_{\infty}\times 90.1\,\frac{\alpha^{6}}{\pi}(\langle V_{\delta,12}\rangle_{v',N'}-\langle V_{\delta,12}\rangle_{v,N})\,\nonumber \\
 & =28.5\,{\rm kHz}\times(\langle V_{\delta,12}\rangle_{v',N'}-\langle V_{\delta,12}\rangle_{v,N})\,.\label{eq:u(delta_f_QED2)}
\end{align}

A smaller uncertainty in $\delta f_{\rm QED}$ arises from the fact that some of the QED corrections at orders $R_{\infty} \alpha^4$ and $R_{\infty} \alpha^5$ have been computed in the adiabatic approximation. This is the case for the relativistic correction of order $R_{\infty} \alpha^4$~\citep{Korobov2017a} and for the one-loop self-energy and vacuum polarization at order $R_{\infty} \alpha^5$~\citep{Korobov2014,Karr2017}. The corresponding uncertainties are estimated from the relative difference between the expectation values of an operator of the type $V_{\delta,12}$ calculated in the adiabatic approximation on the one hand, and in an exact three-body approach on the other hand. These uncertainties are not proportional to $\langle V_{\delta,12}\rangle_{v',N'}-\langle V_{\delta,12}\rangle_{v,N}$, moreover they have very different dependencies on the ro-vibrational degrees of freedom: they are small for transitions between low-lying states and increase when more excited states are involved, whereas $\langle V_{\delta,12}\rangle_{v,N}$ decreases with $v$ and $N$. Due to this, the theoretical errors affecting different transitions are only imperfectly correlated. However, even for transitions between high-lying states, uncertainties associated with the adiabatic approximation remain much smaller than those from uncalculated higher-order terms (Eqs.~\ref{eq:u(delta_f_QED1)} and~\ref{eq:u(delta_f_QED2)}), so that the theoretical uncertainties of all the transitions considered in this work are strongly correlated to each other (see Table~\ref{tab:correl}). This is why in Sec.~\ref{sec:Analytical Model} we neglect uncertainties due to the adiabatic approximation, allowing unknown theoretical contributions to be described by a single unknown parameter. This simplified model yields analytical formulas for the uncertainties of fundamental constants as determined from HMI spectroscopy, which is very useful to guide the choice of an optimal set of transitions. The results of this model are verified by performing a least-squares adjustment (LSA) where imperfect correlations between theoretical uncertainties are taken into account, see Sec.~\ref{sec:LSA}.

There are two other smaller sources of uncertainty from unknown QED contributions. One is a yet uncalculated part of the recoil correction of order $R_{\infty} \alpha^4 (m_e/m_i)$, with estimated uncertainty~\cite{Korobov2017a}
\begin{align}
u(\delta f_{{\rm QED},3}) & =2cR_{\infty}\times \alpha^4 \frac{\pi}{16} \left( \left( \langle V_{\delta,1}\rangle_{v',N'} \!-\!\langle V_{\delta,1}\rangle_{v,N} \right)\frac{m_e}{m_1} + \left( \langle V_{\delta,2}\rangle_{v',N'}\!-\!\langle V_{\delta,2}\rangle_{v,N} \right)\frac{m_e}{m_2}\right)\,.
\end{align}
This can be reexpressed in terms of $\langle V_{\delta,12} \rangle$ and $\langle \Delta V_{\delta,12} \rangle$ similarly to the nuclear-size shift (Eq.~(\ref{delta_f_nuc})). Again, the term proportional to $\langle \Delta V_{\delta,12} \rangle$ is either zero for homonuclear ions, or much smaller for heteronuclear ones, and may be neglected. The uncertainty then simplifies to
\begin{align}
u(\delta f_{{\rm QED},3}) & =2cR_{\infty}\times \alpha^4 \frac{m_e}{\mu_{12}} \frac{\pi}{32} \left( \langle V_{\delta,12}\rangle_{v',N'}-\langle V_{\delta,12}\rangle_{v,N} \right) \nonumber \\
 & =1.5\,{\rm kHz}\times(\langle V_{\delta,pd}\rangle_{v',N'}-\langle V_{\delta,pd}\rangle_{v,N})\,\,\,\,\,\,\,[{\rm HD}^{+}], \\
 & =2.0\,{\rm kHz}\times(\langle V_{\delta,pp}\rangle_{v',N'}-\langle V_{\delta,pp}\rangle_{v,N})\,\,\,\,\,\,\,[{\rm H}_{2}^{+}],
\label{eq:u(delta_f_QED3)}
\end{align}
where $\mu_{12}=m_{1}m_{2}/(m_{1}+m_{2})$.

Finally, the higher-order nuclear correction $\delta f_{\rm nuc,h.o.}$ (last term in eq.~(\ref{eq:f_theor})) is negligibly small for the proton. Only the corrections for the deuteron are included in the theoretical transition frequency, with a small associated uncertainty of~\citep{Korobov2021}
\begin{equation}
   u(\delta f_{\rm nuc,h.o.}) = 0.45 \, {\rm kHz}\times \left( \langle V_{\delta,d} \rangle_{v',N'} - \langle V_{\delta,d} \rangle_{v,N} \right) \, \simeq 0.23 \, {\rm kHz}\times \left( \langle V_{\delta,12} \rangle_{v',N'} - \langle V_{\delta,12} \rangle_{v,N} \right).
   \label{eq:u(delta_f_nuc_ho)}
\end{equation}

The combined uncertainty of the contributions given in eqs.~(\ref{eq: u(delta f_nuc)},\ref{eq:u(delta_f_QED1)},\ref{eq:u(delta_f_QED2)},\ref{eq:u(delta_f_QED3)},\ref{eq:u(delta_f_nuc_ho)})
is 
\begin{eqnarray}
    u_{0}(k)[\langle V_{\delta,12}\rangle_{v',N'}-\langle V_{\delta,12}\rangle_{v,N}]_k\ ,
\end{eqnarray}
where $u_{0}({\rm HD}^{+})\simeq52.2\,{\rm kHz}$ and $u_{0}({\rm H}_{2}^{+})\simeq52.3\,{\rm kHz}$, which includes a 49.9~kHz contribution from purely theoretical uncertainties, and a 15.6~kHz contribution due to the uncertainty of nuclear charge radii in HD$^+$ (15.7~kHz in H$_2^+$). For the vibrational
transitions in Tab.\,\ref{tab:Properties of transitions} the purely theoretical part corresponds
to a fractional frequency uncertainty of approximately
$8\times10^{-12}$. This is a main result of the theory in ref.~\citep{Korobov2021}.
Note that the value of the theoretical uncertainty has no influence on results deduced from the analytical model (see next Section), but is important for the approach based on an LSA (Sec.~\ref{sec:LSA})

One last point related to theoretical uncertainties is worth mentioning. Whereas all contributions to theoretical transition frequencies have been so far computed with negligibly small numerical uncertainties, this may be much harder to achieve for some of the transitions studied in this work, which involve levels with high vibrational quantum numbers such as $v = 18$ (see Tables~\ref{tab:Properties of transitions} and~\ref{tab:Properties of transitions-H2plus}). In particular, accurate calculation of the nonrelativistic Bethe logarithm~\citep{Korobov2012} for such states is a very serious numerical challenge. However, this obstacle is not of a fundamental nature, and in the present exploratory analysis we will assume that it can be overcome in the future.

\section{Analytical model}
\label{sec:Analytical Model}
\subsection{Master equation}
Above, we have clarified that the combined uncertainty of the transition
frequency arising from nuclear and QED effects is, to a good approximation,
proportional to the delta-potential difference of lower and upper
spectroscopy levels. This is the key point for our analytical model. We shall
recast this uncertainty as an unknown correction to be determined
from experiments. To quantify the correction we introduce the species-dependent, dimensionless
parameter $\Delta_{\rm nuc,QED}(k)$ that
describes the combined deviations from the ``best'' theory values,
\[
\Delta_{\rm nuc,QED}(k)[\langle V_{\delta,12}\rangle_{v',N'}-\langle V_{\delta,12}\rangle_{v,N}]_{k}\times2cR_{\infty}\alpha^{5}=\Delta(\delta f_{\rm nuc}+\delta f_{\rm nuc,h.o.}+\delta f_{{\rm QED},1}+\delta f_{{\rm QED},2}+\delta f_{{\rm QED},3})\ ,
\]
where $k$ denotes the molecular species. $\Delta \delta f_i$ is the unknown deviation of the actual contribution of type $i$ from the currently calculable one. 

According to the discussion above, the recoil correction ($\delta f_{{\rm QED},3})$ is species dependent. There are different possibilities for its treatment. The correction is amenable to an \emph{ab initio} calculation~\citep{Zhong2018}, and this calculation is expected to be easier than reducing $u(\delta f_{\rm QED,1})$ or $u(\delta f_{\rm QED,2})$. In this case, we would be allowed to ignore the correction in the present context. Alternatively, we could incorporate the recoil correction into $\Delta_{\rm nuc}(k)$. With any of these treatments, we may write
\[
\Delta_{\rm nuc,QED}(k)=\Delta_{\rm nuc}(k)+\Delta_{\rm nuc,h.o.}(k)+\Delta_{\rm QED}\,\,,
\]
where all quantities are dimensionless and independent of the levels,
and $\Delta_{\rm QED}$ is also independent of the molecule species.

The Rydberg constant is also affected by uncertainty; we introduce
the fractional deviation with respect to the CODATA 2018 value $\Delta_{h}=R_{\infty}/R_{\infty,2018}-1$.

In summary, we may express a theoretical spin-averaged frequency as
follows,
\begin{align}
f^{\rm (theor)}(v,N,v',N') & =f_{{\rm 2018}}^{\rm (theor)}(v,N,v',N')+2c\,R_{\infty,2018}\times\nonumber \\
 & \phantom{=}\Bigr(\Delta_{h}f_{{\rm 2018},at.u.}^{\rm (theor)}(v,N,v',N')+\nonumber \\
 & \phantom{=\Bigr(}\sum_{i=1,2}\Delta_{m,i}\,\partial f_{{\rm 2018},at.u.}^{\rm (theor)}(v,N,v',N')/\partial(m_{i}/m_{e})+\nonumber \\
 & \phantom{=\Bigr(}\Delta_{\rm nuc,QED}\alpha^{5}(\langle V_{\delta,12}\rangle_{v',N'}-\langle V_{\delta,12}\rangle_{v,N})\Bigl)\,,\label{eq:master equation}
\end{align}
where $\Delta_{m,i}=m_{i}/m_{e}-[m_{i}/m_{e}]_{{\rm 2018}}$ and
the molecule-species dependence $k$ is implicit. $f_{{\rm 2018}}^{\rm (theor)}(v,N,v',N')$
is the \emph{ab initio} transition frequency computed with the CODATA~2018 fundamental constants and no uncertainty is associated with it.
A subscript ``at.u.'' indicates that the frequency is expressed in
atomic units. 
The above ``master equation''
is a generalization of eq.~(16) of ref.~\citep{Korobov2017a}. (While
the master equation has been stated with CODATA 2018 values, these
are just reference values; the equation would be just as applicable
if we used CODATA~2014 reference values instead.) There is one such
master equation for each MHI species. For homonuclear MHI appropriate
simplifications hold. Considering all six MHI species (enumerated
by the index $k$), there are 10 unknown parameters, $\Delta_{m,p}$,
$\Delta_{m,d}$, $\Delta_{m,t}$, $\{\Delta_{\rm nuc,QED}(k)\}$, and
$\Delta_{h}$.

The master equation is approximate, since we have explicitly neglected
some terms that arise for heteronuclear MHI and because additional
small uncertainties are present in the theoretical transition frequency,
that stem from the use of adiabatic wavefunction to calculate several QED contributions and do not have the simple form of the last term in eq.\,(\ref{eq:master equation}).

For exploratory analyses such as the present one, in the second and third line of eq.\,(\ref{eq:master equation}),
$f_{{\rm 2018}}^{\rm (theor)}$ may be replaced by $f_{nr}$, the
theoretical nonrelativistic frequency. The latter can be approximated by the adiabatic transition frequency
- obtained by solving the one-dimensional radial Schr\"odinger equation
with the adiabatic potential \citep{Korobov2018} appropriate to each
species.

For heteronuclear MHI, the two mass-deviation contributions from $m_1$ and $m_2$ may be
approximately subsumed into a single one concerning the reduced nuclear
mass $\mu_{12}$:
\[
\Delta_{m,\lambda}\,\frac{\partial f_{nr,at.u.}(v,N,v',N')}{\partial \lambda}\,,
\]
with $\lambda=\mu_{12}/m_{e}$, $\Delta_{m,\lambda}=\mu_{12}/m_{e}-[\mu_{12}/m_{e}]_{{\rm 2018}}$.
This approximation is good, because the nonrelativistic transition
frequency is well approximated by the difference of the adiabatic
energies.

For reference, the current (CODATA 2018) uncertainties of the mass ratio deviations
are  $u(\Delta_{m,p,2018})=1.1\times10^{-7}$,  $u(\Delta_{m,d,2018})=1.3\times10^{-7}$,
$u(\Delta_{m,t,2018})=2.7\times10^{-7}$. Furthermore
the (current) uncertainty of $\Delta_{\rm nuc,QED}$ is $u(\Delta_{\rm nuc,QED}({\rm HD}^{+}))\simeq52\,{\rm kHz}/(2\,c\,R_{\infty}\alpha^{5})\simeq0.38$,
and similar for H$_{2}^{+}$. 
The uncertainty of $\lambda=\mu_{pd}/m_{e}$
is $u([\Delta_{m,\lambda}({\rm HD}^{+})]_{2018})=5.6\times10^{-8}$,
or $4.6\times10^{-11}$ in relative terms. 
The uncertainty of this last quantity deduced
from recent Penning trap experiments~\citep{Koehler2015,Heisse2019,Rau2020,Fink2021} is 
moderately smaller~\cite{Karr2023},
see Fig.~\ref{fig:Four-HD-transitions}. All of these uncertainties do not enter the
present treatment. 

The Rydberg constant fractional uncertainty is
$u(\Delta_{h,2018})=$ $1.9\times10^{-12}$. This is a relevant quantity only in the
simplest scenario outlined here, the measurement of a transition pair 
(Supplemental Material (SM), Tab.\,I).

\vfill
\newpage
\subsection{Transitions}

Table~\ref{tab:Properties of transitions} presents the relevant
parameters for a number of electric-dipole allowed transitions of HD$^+$.
A few electric-quadrupole transitions of H$_{2}^{+}$ having $v'-v=2$
are reported in Tab.~\ref{tab:Properties of transitions-H2plus}.

The mass sensitivities, denoted by $s$ or $\beta_\lambda$ in the following sections, as well as the expectation values of the delta functions, $\langle V_\delta\rangle$, are obtained from nonadiabatic (exact) nonrelativistic calculations, a choice which may be relevant especially for transitions involving large vibrational quantum numbers.
To compute the mass sensitivities we followed an approach similar to that of ref.~\citep{Schiller2005}. The hamiltonian (Eq.~(6) of~\citep{Schiller2005}) is re-expressed as a function of the variables $\mu_{pd}/m_e$ and $m_p/m_e$, then the derivatives are easily found to be equal to some combinations of the kinetic energy operator expectation values. Their expressions can be found in eqs.\,(13\,b) and (13\,c) of ref.~\citep{Karr2023}.

Among the shown transitions of HD$^+$ are those
that have already been studied experimentally to date, all having
$v=0$ as lower level. In addition, a few hot-band transitions are
included. Transition number\,5 (red in Fig.~1) occurs between highly
excited vibrational levels: $v=9$, $v'=18$. It exhibits the opposite
sign of the sensitivity $s$ of the frequency to the mass ratio compared
to transitions having the ground vibrational level as lower level.
Transition 5 is just one of several such transitions. Also, we point
out the existence of transitions that have strongly suppressed sensitivity
to the mass ratio. Number 6 is an example that is shown in blue in
Fig.~1. Such transitions are obviously not effective in determining
the mass ratio. However, they are effective in determining the QED
contribution parameter $\Delta_{\rm nuc,QED}$, and therefore, when part
of a set of transitions, contribute to determining the mass
ratio.
An important aspect is that while $s$ does not scale with the transition
frequency value, $V$ is approximately proportional to it. This correlation
has an important consequence: $\Delta_{\rm nuc,QED}$ cannot be determined
as accurately as the mass ratio.

\begin{table}[t]
$$\begin{array}{|c|c|c|c|c|c|c|c|}
\hline
 \text{no.} & \text{Transition} & \text{Frequency} & f_{\text{nr},\text{at.u.}} & s & V & \text{$\Delta V$} \\
  & (v,N)\rightarrow (v',N') & f\,\text{(THz)} &  & \partial f_{\text{nr},\text{at.u.}}\text{/$\partial $}\lambda & \left\langle V_{\delta }\rangle _{v',N'}\text{-$\langle $}V_{\delta }\rangle _{v,N}\right. & \left\langle \text{$\Delta V$}_{\delta }\rangle _{v',N'}\text{-$\langle $}\text{$\Delta V$}_{\delta }\rangle _{v,N}\right. \\
\hline
 1 & (0,0\text{)$\to $(}0,1) & 1.31 & 0.000200 & -1.611\times 10^{-7} & -0.0003698 & 2.0\times10^{-10} \\
\hline
 2 & (0,0\text{)$\to $(}1,1) & 58.6 & 0.00891 & -3.526\times 10^{-6} & -0.009850 & -6.8\times10^{-6} \\
\hline
 3 & (0,0\text{)$\to $(}5,1) & 260 & 0.0395 & -1.363\times 10^{-5} & -0.04187 & -0.000057 \\
\hline
 4 & (0,3\text{)$\to $(}9,3) & 415 & 0.0631 & -1.816\times 10^{-5} & -0.06460 & -0.00018 \\
\hline
 5 & (9,1\text{)$\to $(}18,0) & 208 & 0.0316 & 8.272\times 10^{-6} & -0.02868 & -0.0043 \\
\hline
 6 & (7,1\text{)$\to $(}15,0) & 237 & 0.0360 & -3.113\times 10^{-8} & -0.03373 & -0.00095 \\
\hline
 7 & (9,1\text{)$\to $(}13,0) & 118 & 0.0180 & -1.027\times 10^{-7} & -0.01671 & -0.00036 \\
\hline
 8 & (5,1\text{)$\to $(}13,0) & 278 & 0.0422 & -5.242\times 10^{-6} & -0.04059 & -0.00049 \\
\hline
\end{array}
$$
\caption{\label{tab:Properties of transitions}
Some transitions of HD$^{+}$
and their properties. The first one is a rotational transition,
the others are vibrational transitions. The last four entries are
\textquotedblleft hot-band\textquotedblright{} transitions. Transition
5 is the only one among the list that has a positive $s$. $\lambda=\mu_{pd}/m_e$. In this table, $V_\delta\equiv V_{\delta,pd}$,  $\Delta V_\delta\equiv \Delta V_{\delta,pd}$. 
}
\end{table}
\begin{table}[h]
$\begin{array}{|c|c|c|c|c|c|c|}
\hline
 \text{no.} & \text{Transition} & \text{Frequency} & f_{\text{nr},\text{at.u.}} & s & V \\
  & (v,N)\to (v',N') & f\text{(THz)} &  & \partial f_{\text{nr},\text{at.u.}}\text{/$\partial $(}m_p/m_e) & \left\langle V_{\delta }\rangle _{v',N'}\text{-$\langle $}V_{\delta }\rangle _{v,N}\right. \\
\hline
 1 & (1,0\text{)$\to $(}3,2) & 124 & 0.0189 & -4.531\times 10^{-6} & -0.02049 \\
\hline
 2 & (11,0\text{)$\to $(}13,2) & 53.4 & 0.00811 & 1.516\times 10^{-6} & -0.007464 \\
\hline
 3 & (12,0\text{)$\to $(}14,2) & 45.9 & 0.00698 & 2.284\times 10^{-6} & -0.006314 \\
\hline
 4 & (9,0\text{)$\to $(}11,2) & 67.7 & 0.0103 & 1.572\times 10^{-7} & -0.009773 \\
\hline
\end{array}
$
\caption{\label{tab:Properties of transitions-H2plus}
Some rovibrational transitions
of H$_{2}^{+}$ and their properties. The first transition has recently
been observed \citep{Schenkel2023}. Transitions 2 and 3 have substantial positive
sensitivity $s$. Transition 4 has a suppressed sensitivity. In this table, $V_\delta\equiv V_{\delta,pp}$.}
\end{table}

\vfill
\newpage
\subsection{Determination of the mass ratios and Rydberg constant}

If a large enough set of experimental transition frequencies is available,
a LSA of the quantities $\Delta_{m,\mu}$, $\Delta_{\rm nuc,QED}(k)$
and $\Delta_{h}$ can be made using the master equation~(\ref{eq:master equation}).
Instead, for the sake of simplicity we consider the cases of minimal-size
data sets being available, so that these quantities are obtained by solving a linear system of equations. 

The proposed measurement approach presented here is not only applicable
to HD\textsuperscript{+} but also to all other MHI species - whose
vibrational transitions have not yet been determined with laser spectroscopy
with competitive accuracy. 
We have analyzed several scenarios. Some of them are presented in SM. One is described in the following.

In view of the approximations made, projected uncertainty levels obtained with the analytical approach 
should be considered as indicative only. The analytical approach is nevertheless very useful as it allows to identify easily the most promising transitions, as confirmed by comparison with an LSA in Sec.\,\ref{sec:LSA}.

\subsubsection*{Two species}
One scenario we consider 
comprises
three transitions
in HD$^{+}$ and two transitions in H$_{2}^{+}$.
This will allow to determine the five quantities $\mu_{pd}/m_{e}$, $m_{p}/m_{e}$,
$\Delta_{\rm nuc,QED}({\rm H}_{2}^{+})$, $\Delta_{\rm nuc,QED}({\rm HD}^{+})$,
and $R_{\infty}$, that appear in the two corresponding master equations.
Once $\mu_{pd}/m_{e}$ and $m_{p}/m_{e}$ have been obtained, the
deuteron mass $m_{d}/m_{e}$ becomes available. Data from more than
5 transitions would be very useful for consistency checks, and would
then be analyzed using LSA.

Let us first consider a general aspect of this scenario. Recall the
deviations $\Delta_{\rm nuc,QED}({\rm H}_{2}^{+})$, $\Delta_{\rm nuc,QED}({\rm HD}^{+})$
in terms of their contributions, the charge radii deviations, the
QED deviation, and the higher-order nuclear deviation (if present),
\begin{align}
\Delta_{\rm nuc,QED}({\rm H}_{2}^{+}) & =\alpha^{-5}(2\pi/3)a_{0}^{-2}\frac{1}{2}\times2\,\Delta(r_{p}^{2})+\Delta_{\rm QED}\,,\label{eq:Delta_nuc,QED for H2+}\\
\Delta_{\rm nuc,QED}({\rm HD}^{+}) & =\alpha^{-5}(2\pi/3)a_{0}^{-2}\frac{1}{2}\times\Delta(r_{p}^{2}+r_{d}^{2})+\Delta_{\rm QED}+\Delta{}_{\rm nuc,h.o.}(d)\,.\label{eq:Delta_nuc,QED(HD+)}
\end{align}
Small corrections have been neglected. Note that $\Delta_{\rm QED}$ is
independent of the molecular species, under the assumptions made. 
We may compare the contributions on each r.h.s., in other words, their  uncertainties. The uncertainty of the first contribution is of order $0.11$.
The {uncertainty of the} second, {$u(\Delta_{\rm QED})$}, is of order $50\,\alpha\approx0.37$, according to eqs.~(\ref{eq:u(delta_f_QED1)},\ref{eq:u(delta_f_QED2)}).
Last, $u(\Delta{}_{\rm nuc,h.o.}(d)) \simeq$ 
$(0.23\,{\rm kHz})/(2cR_{\infty}\alpha^{5})=0.0017$
(eq.~(\ref{eq:u(delta_f_nuc_ho)})). 
Once fit results for $\Delta_{\rm nuc,QED}({\rm H}_{2}^{+})$, $\Delta_{\rm nuc,QED}({\rm HD}^{+})$ are obtained,
we can obtain from eqs.\,(\ref{eq:Delta_nuc,QED for H2+},\ref{eq:Delta_nuc,QED(HD+)}) an approximate value for the difference of the squared
radii of deuteron and proton,
\begin{align}
\Delta(r_{d}^{2})-\Delta(r_{p}^{2})\simeq2(2\pi/3)^{-1}a_{0}^{2}\,\alpha^{5}\left(\Delta_{\rm nuc,QED}({\rm HD}^{+})-\Delta_{\rm nuc,QED}({\rm H}_{2}^{+})-\Delta{}_{\rm nuc,h.o.}(d)\right)\,.
\label{eq:rd2_minus_rp2_from_H2plus_and_HDplus}
\end{align}
The {mentioned} uncertainty of the last term sets the minimum possible uncertainty
of the l.h.s., $9\times10^{-5}\,{\rm fm}^{2}$. This is a factor 4.5
less than the CODATA 2018 uncertainty, eq.~(\ref{eq:rd2-rp2}). 

Table~\ref{tab:Five transitions} shows the result of a particular
measurement scenario: three ${\rm HD}^{+}$ transitions found 
to be favourable (see SM) and two transitions in ${\rm H}_{2}^{+}$. 

For frequency uncertainties at the 1\,Hz level (last line in the
table), the uncertainty of {the Rydberg constant is one order smaller than CODATA 2018, while for $\mu_{pd}/m_e$ it is two orders and for $m_p/m_e$ it is three orders smaller.}
The uncertainty of $\Delta_{\rm nuc,QED}({\rm HD}^{+})-\Delta_{\rm nuc,QED}({\rm H}_{2}^{+})$
is smaller than that of $\Delta_{\rm nuc,h.o.}(d)$, see rightmost column in the table. This implies an experimental uncertainty for
$\Delta(r_{d}^{2})-\Delta(r_{p}^{2})$  four times smaller than the CODATA
2018 uncertainty. 
{These results are encouraging and will be verified in the next chapter using a more rigorous approach.}

\begin{table}[h]
\footnotesize{
$
\begin{array}{|c|c|c|c|c|c|c|c|c|c|c|c|}
\hline

\multicolumn{5}{|c|} {\text{Transitions}} & u_a,\,u_b,\,u_c,\,u_d,\,u_e & u\left(\Delta _{m,\lambda }\right) & u\left(\Delta _{m,p}\right) & u_r(R_\infty)
& 
u(\Delta_{{\rm HD}^+})
& 
u(\Delta_{{\rm H}_2^+})
& 
u(\Delta_{{\rm HD}^+}-
\\

 \text{a} & \text{b} & \text{c} & \text{d} & \text{e} & \text{(kHz)} & \left(10^{-10}\right) & \left(10^{-10}\right) & \left(10^{-12}\right) & \text{} & \text{} 
 & 
 \Delta_{{\rm H}_2^+})
 \\
\hline
 3 & 4 & 5 & 1 & 2 & \{0.3,0.3,0.3,0.3,0.1\} & 130 & 220 & 55 & 2.7 & 2.7 & 0.11 \\
\hline
 3 & 4 & 5 & 1 & 2 & \{0.1,0.1,0.1,0.1,0.03\} & 44 & 74 & 18 & 0.91 & 0.89 & 0.037 \\
\hline
 3 & 4 & 5 & 1 & 2 & \{0.03,0.03,0.03,0.03,0.03\} & 13 & 26 & 5.5 & 0.27 & 0.27 & 0.018 \\
\hline
 3 & 4 & 5 & 1 & 2 & \{0.003,0.003,0.003,0.01,0.01\} & 1.3 & 5.5 & 0.55 & 0.027 & 0.027 & 0.0055 \\
\hline
 3 & 4 & 5 & 1 & 2 & \{0.003,0.003,0.003,0.003,0.003\} & 1.3 & 2.6 & 0.55 & 0.027 & 0.027 & 0.0018 \\
\hline
 3 & 4 & 5 & 1 & 2 & \{0.003,0.003,0.003,0.001,0.001\} & 1.3 & 2.2 & 0.55 & 0.027 & 0.027 & 0.0010 \\
\hline
 3 & 4 & 5 & 1 & 2 & \{0.001,0.001,0.001,0.001,0.001\} & 0.44 & 0.87 & 0.18 & 0.0091 & 0.0089 & 0.00061 \\
\hline
\end{array}
$

}
\caption{\label{tab:Five transitions}
Analytical model: determination of the five quantities 
$m_{p}/m_{e}$, 
$\lambda=\mu_{pd}/m_{e}$,
Rydberg constant, and nuclear-QED corrections $\Delta_{\rm nuc,QED}({\rm H}_{2}^{+})$,
$\Delta_{\rm nuc,QED}({\rm HD}^{+})$ by measuring five transitions,
three in HD\protect\textsuperscript{+}  and two in ${\rm H_{2}^{+}}$.  Columns 1,~2,~3 indicate the chosen transitions of HD\protect\textsuperscript{+} (a,b,c);
columns~4, 5 refer to the transitions of ${\rm H_{2}^{+}}$ (d,e). The transition labels are
defined in previous tables. Column~6 are the assumed experimental uncertainties. The other columns give the absolute uncertainties
of the determined quantities. Note that the fractional uncertainty of the Rydberg constant, $u_r(R_\infty)$, is equal to $u(\Delta_{h})$. We used the abbreviations 
$\Delta_{{\rm HD}^+}=\Delta _{\text{nuc},\text{QED}}(\text{HD}^+)$, 
$\Delta_{{\rm H}_2^+}=\Delta _{\text{nuc},\text{QED}}(\text{H}_2^+)$.
}
\end{table}

\vfill
.\newpage
\eject
.\eject
\section{Simulations of fundamental constant determinations using
least-squares adjustments} \label{sec:LSA}

The analysis presented in Sec.\,\ref{sec:Analytical Model} relies on the assumption that the combined uncertainty of a transition frequency arising from nuclear and QED effects can be described by a single term proportional to the delta-potential difference between the lower and upper levels. As already noted, this is only approximately true, because one of the contributions to the theoretical uncertainty cannot be written under this form, namely, the uncertainty arising from the use of the adiabatic approximation in calculation of high-order QED corrections. A more elaborate description that takes into account the imperfect correlations between theoretical uncertainties of different rovibrational transition frequencies is thus required to verify the insights given by our simple analytical model and obtain more precise estimates of achievable precision of fundamental constant determinations from MHI spectroscopy. In the following, we use the linearized least-squares adjustment (LSA) procedure described in Appendix E of~\cite{Mohr2000}. See also~\cite{Karr2023} for a recent application to MHI spectroscopic data.

Each of the (hypothetical) transition frequency measurements yields the following observational equations:
\begin{eqnarray} \label{eq:obseq1}
f^{\rm exp} &\doteq& f^{\rm th} + \delta f^{\rm th} + \beta_{\lambda} (\lambda \!-\! \lambda_0) + \beta_{R_{\infty}} c(R_{\infty} \!-\! R_{\infty 0}) + \beta_{r_i} (r_i \!-\! r_{i0})\,, \\
\label{eq:obseq2}
\delta f &\doteq& \delta f^{\rm th}\ .
\end{eqnarray}
The dotted equality sign means that the left and right hand sides should agree within estimated uncertainties. $f^{\rm exp}$ and $f^{\rm th}$ are respectively the experimental and theoretical transition frequency. The latter is obtained using reference (e.g. CODATA 2018) values of the involved fundamental constants: the mass ratio $\lambda$, the Rydberg constant, and the nuclear charge radii $r_i$. The dependence of the theoretical frequency on these constants is linearized around their reference values ($\lambda_0$, $R_{\infty 0}$, $r_{i0}$) using the sensitivity coefficients: $\beta_{\lambda} = \partial f^{\rm th}/\partial \lambda \simeq \partial f^{\rm th}_{\rm nr}/\partial \lambda$, $\beta_{R_{\infty}} = f^{\rm th}/(cR_{\infty})$, $\beta_{r_i} = \partial f^{\rm th}/\partial r_i$. For heteronuclear molecules, eq.~(\ref{eq:obseq1}) contains an implicit summation over $i$, and $\lambda \equiv \mu_{12}/m_e$, $\mu_{12}$ being the reduced mass of the nuclei. For homonuclear molecules, $\lambda \equiv m_i/m_e$, $m_{i}$ being the mass of a nucleus. 

The theoretical uncertainty of the transition frequency is accounted for by introducing the additive correction $\delta f^{\rm th}$, which is treated as an adjusted constant. A second input data with zero value ($\delta f \equiv 0$) and uncertainty equal to the estimated theoretical uncertainty is included in the LSA [Eq.~(\ref{eq:obseq2})].

We take into account correlations between different input data. Measurements of different transitions are assumed to be uncorrelated to each other, but theoretical uncertainties are strongly correlated. The correlation coefficients among the  $\delta f$ are estimated using the results of~\cite{Korobov2021} and can be found in Table~\ref{tab:Correl-th}.

\begin{table}
	\begin{tabular}{@{\hspace{2mm}}l@{\hspace{3mm}}l@{\hspace{3mm}}l@{\hspace{3mm}}l@{\hspace{3mm}}l@{\hspace{2mm}}} 
		\hline
		\multicolumn{5}{c}{Correlation coefficients} \\
		\hline \hline
        \multicolumn{5}{c}{Among HD$^+$ transitions} \\
        \hline
		$r(1,2) = 0.99570$ & $r(1,3) = 0.98071$ & $r(1,4) = 0.95733$ & $r(1,5) = 0.91011$ & $r(1,6) = 0.86385$ \\
        $r(1,7) = 0.84570$ & $r(1,8) = 0.87967$ & $r(2,3) = 0.99460$ & $r(2,4) = 0.97998$ & $r(2,5) = 0.94457$ \\
        $r(2,6) = 0.90678$ & $r(2,7) = 0.89148$ & $r(2,8) = 0.91993$ & $r(3,4) = 0.99535$ & $r(3,5) = 0.97355$ \\
        $r(3,6) = 0.94565$ & $r(3,7) = 0.93370$ & $r(3,8) = 0.95566$ & $r(4,5) = 0.99102$ & $r(4,6) = 0.97257$ \\
        $r(4,7) = 0.96383$ & $r(4,8) = 0.97958$ & $r(5,6) = 0.99493$ & $r(5,7) = 0.99081$ & $r(5,8) = 0.99766$ \\
        $r(6,7) = 0.99939$ & $r(6,8) = 0.99948$ & $r(7,8) = 0.99774$ & & \\        
		\hline
        \multicolumn{5}{c}{Among H$_2^+$ transitions} \\
        \hline
	    $r(1,2) = 0.93922$ & $r(1,3) = 0.93365$ & $r(1,4) = 0.92866$ &
        $r(2,3) = 0.99987$ & $r(2,4) = 0.99956$ \\
        $r(3,4) = 0.99991$ & & & & \\
        \hline
        \multicolumn{5}{c}{Between HD$^+$ and H$_2^+$ transitions} \\
        \hline
        $r(1,1) = 0.99083$ & $r(1,2) = 0.88436$ & $r(1,3) = 0.87684$ & $r(1,4) = 0.87018$ & $r(2,1) = 0.99904$ \\
        $r(2,2) = 0.92377$ & $r(2,3) = 0.91758$ & $r(2,4) = 0.91206$ & $r(3,1) = 0.99804$ & $r(3,2) = 0.95852$ \\
        $r(3,3) = 0.95388$ & $r(3,4) = 0.94969$ & $r(4,1) = 0.98748$ & $r(4,2) = 0.98149$ & $r(4,3) = 0.97833$ \\
        $r(4,4) = 0.97541$ & $r(5,1) = 0.95759$ & $r(5,2) = 0.99826$ &
        $r(5,3) = 0.99721$ & $r(5,4) = 0.99610$ \\
        $r(6,1) = 0.92378$ & $r(6,2) = 0.99905$ & $r(6,3) = 0.99960$ & $r(6,4) = 0.99987$ & $r(7,1) = 0.90982$ \\
        $r(7,2) = 0.99696$ & $r(7,3) = 0.99806$ & $r(7,4) = 0.99881$ &
        $r(8,1) = 0.93566$ & $r(8,2) = 0.99990$ \\
        $r(8,3) = 0.99993$ & $r(8,4) = 0.99976$ & & & \\
        \hline
	\end{tabular}
	\caption{Correlation coefficients between theoretical uncertainties of HD$^+$ and H$_2^+$ transition frequencies. The arguments are the transition numbers, defined in Tables \ref{tab:Properties of transitions},\,\ref{tab:Properties of transitions-H2plus}. \label{tab:correl} \label{tab:Correl-th}}
\end{table}

\newpage
\subsubsection{Two transitions in one species}
\label{sec:LSA two transitions in one species}
We first consider the scenario where two transitions are measured in HD$^+$. Since in this case we are only aiming  to determine the mass ratio $\mu_{pd}/m_e$, 
additional data is required on the other involved fundamental constants: $R_{\infty}$, $r_p$ and $r_d$. We thus include the following observational equations:
\begin{eqnarray}
R_{\infty,2018} &\doteq& R_{\infty}, \\
r_{i,2018} &\doteq& r_i \;\; (i = p,d).
\end{eqnarray}
The correlation coefficients between the CODATA 2018 values of these constants are: $r(R_{\infty,2018}, r_{p,2018}) = 0.88592$, $r(R_{\infty,2018}, r_{d,2018}) = 0.90366$, and $r(r_{p,2018}, r_{d,2018}) = 0.99165$. The total number of input data is $N=7$ (two experimental frequencies, two associated $\delta f$, and the three CODATA 2018 values), whereas the number of adjusted constants is $M=6$ ($\mu_{pd}/m_e$, $R_{\infty}$, $r_p$, $r_d$, and the two $\delta f^{\rm th}$). Results are displayed in the last column of Tab.\,I in SM.

As already indicated by the analytical model (see SM), we recognize the importance of choice of transition pair: (3,5) is substantially more favourable than (3,4). The lowest LSA uncertainty among the examples is approximately $3\times10^{-9}$, a factor~8 lower than what is obtained from two of the currently available measurements (last row in the table), and a factor 19 smaller than the CODATA 2018 uncertainty.

In the case of H$_2^+$, two measurements may already offer the possibility to determine other constants in addition to $m_p/m_e$. However, two transitions are still not sufficient to determine the three involved constants ($m_p/m_e$, $R_{\infty}$, $r_p$), and some additional input data is required. It would make little sense to include the CODATA 2018 value of, e.g., $r_p$ in order to determine $m_p/m_e$ and $R_{\infty}$ because the proton charge radius is strongly correlated to the Rydberg constant by the very precise measurements of the 1S-2S transition frequency in the H atom~\cite{Parthey2011,Matveev2013}. We instead include these 1S-2S measurements and associated theoretical $\delta f$ 
correction (similarly to eq.~(\ref{eq:obseq2}))  in our LSA using the information provided in~\cite{Tiesinga2021} (items A6-A7 from Table X, B1-B2 from Table VIII, and their correlation coefficients from Table IX). The uncertainty of the theoretical correction is 1.4\,kHz. The total number of input data is $N=7$ (two H$_2^+$ experimental frequencies, two associated $\delta f$, the two H(1S-2S) measurements and one associated $\delta f$), whereas the number of adjusted constants is $M=6$ ($m_p/m_e$, $R_{\infty}$, $r_p$, the two $\delta f^{\rm th}$(H$_2^+$) and one $\delta f^{\rm th}$(H)). 

Results are displayed in Tab.~\ref{tab:Two transitions H2+ with LSA}. We see that there is no further gain in pursuing an experimental frequency uncertainty smaller than 0.1\,kHz. For that case, the mass ratio uncertainty, $\simeq~4.7\times10^{-12}$ fractionally, is 13~times smaller than the CODATA 2018 uncertainty.  
\begin{table}[tb!]
$\begin{array}{|c|c|c|c|c|c|}
\hline
\multicolumn{2}{|c|} {\text{Trans.}} 
& u_a,u_b & u_r (m_p/m_e) & u_r (R_{\infty}) & u\left(r_p\right) \\
\text{a} & \text{b} & \text{(kHz)} & \left(10^{-12}\right) & \left(10^{-12}\right) & \text{(fm)} \\
\hline
 1 & 3 & \{0.3,0.3\} & 7.4 & 27 & 0.029 \\
\hline
 1 & 3 & \{0.1,0.1\} & 4.7 & 25 & 0.027 \\
\hline
 1 & 3 & \{0.03,0.03\} & 4.3 & 25 & 0.027 \\
\hline
 1 & 3 & \{0.01,0.01\} & 4.2 & 25 & 0.027 \\
\hline
 1 & 3 & \{0.003,0.003\} & 4.2 & 25 & 0.027 \\
\hline
 1 & 3 & \{0.001,0.001\} & 4.2 & 25 & 0.027 \\
 \hline
 \hline
 \multicolumn{3}{|c|}{\text{CODATA 2018}}& 60 & 1.9 & 0.0019 \\ 
\hline
\end{array}$
\caption{Linearized LSA procedure. Examples of determination of 
$m_p/m_e$, Rydberg constant, and proton charge radius by measuring two transitions a, b in H$_2^+$.  
H(1S-2S) measurements are included in the input data. $u_r$ denotes a fractional uncertainty, $u$ an absolute uncertainty.} \label{tab:Two transitions H2+ with LSA}
\end{table}

\newpage
\subsubsection{Three transitions in one species}
\label{sec:LSA three transitions in one species}

Similarly, three transition measurements in HD$^+$ could be used to determine the Rydberg constant and nuclear radii $r_p$, $r_d$. To do this, we again include the H(1S-2S) measurements~\cite{Parthey2011,Matveev2013}, but also the H-D(1S-2S) isotope shift measurement~\cite{Parthey2010} (item A5 from Table X, B17-B18 from Table VIII, and correlation coefficients from Table IX). The theory uncertainty is set to 0.34\,kHz. The total number of input data is $N=11$ (three HD$^+$ experimental frequencies, three associated $\delta f$, the three H/D(1S-2S) measurements and two associated $\delta f$), whereas the number of adjusted constants is $M=9$ ($\mu_{pd}/m_e$, $R_{\infty}$, $r_p$, $r_d$, the three $\delta f^{\rm (th)}$(HD$^+$) and $\delta f^{\rm (th)}$(H), $\delta f^{\rm (th)}$(H-D)). 

Results are displayed in Tab.~\ref{tab:Three transitions with LSA}. The first case in the table (data row 1) considers the three rovibrational transitions measured to date. In this case, the uncertainty of the fitted mass ratio $\mu_{pd}$ is not competitive because no input values of the Rydberg constant and nuclear radii are provided.
Among the examples, the lowest LSA uncertainty for the mass ratio is approximately 
$3\times10^{-9}$, the same value as for the case of two transitions only. Now, also the charge radii are determined, but their uncertainties are approximately one order larger than from muonic hydrogen/deuterium measurements.

\begin{table}[h!]
\footnotesize{
$\begin{array}{|c|c|c|c|c|c|c|c|}
\hline
\multicolumn{3}{|c|} {\text{Trans.}} & u_a,u_b,u_c & u_r (\lambda_{pd}) & u_r (R_{\infty}) & u\left(r_p\right) & u\left(r_d\right) \\

\text{a} & \text{b} & \text{c} & \text{(kHz)} & \left(10^{-12}\right) & \left(10^{-12}\right) & \text{(fm)} & \text{(fm)} \\

\hline
 2 & 3 & 4 & \{0.15,0.6,0.46\} & 106 & 111 & 0.12 & 0.047 \\
\hline
 3 & 4 & 5 & \{0.3,0.3,0.3\} & 5.4 & 23 & 0.024 & 0.0095 \\
\hline
 3 & 4 & 5 & \{0.03,0.03,0.03\} & 2.2 & 17 & 0.018 & 0.0071 \\
\hline
 3 & 4 & 5 & \{0.01,0.01,0.01\} & 2.1 & 17 & 0.018 & 0.0070 \\
\hline
 3 & 4 & 5 & \{0.003,0.003,0.003\} & 2.1 & 17 & 0.018 & 0.0070 \\
\hline
 3 & 4 & 5 & \{0.001,0.001,0.001\} & 2.0 & 17 & 0.018 & 0.0070 \\
\hline
 3 & 5 & 6 & \{0.003,0.003,0.003\} & 3.0 & 19 & 0.021 & 0.0082 \\
\hline
 3 & 5 & 8 & \{0.003,0.003,0.003\} & 2.9 & 19 & 0.020 & 0.0081 \\
\hline
 3 & 6 & 7 & \{0.003,0.003,0.003\} & 2.4 & 18 & 0.019 & 0.0076 \\
\hline
 3 & 6 & 8 & \{0.003,0.003,0.003\} & 2.6 & 19 & 0.020 & 0.0078 \\
\hline
\hline
\multicolumn{4}{|c|}{\text{CODATA 2018}}& 46 & 1.9 & 0.0019 & 0.00074 \\
\hline
\end{array}$
}
\caption{Examples of LSA 
of the four constants $\lambda_{pd} = \mu_{pd}/m_e$, $R_\infty$, $r_p$, $r_d$ by measuring three transitions a, b, c in HD$^+$.  
H(1S-2S) and H-D(1S-2S) isotope shift measurements are included in the input data.} 
\label{tab:Three transitions with LSA}
\end{table}

\newpage

\subsubsection{Two species}

Here, we consider the case where three transition measurements in HD$^+$ and two in H$_2^+$ are available. In principle, this is enough to determine the five quantities: $\mu_{pd}/m_e$, $m_p/m_e$, Rydberg constant, proton and deuteron charge radii. The LSA then comprises $N=10$ input data (the five experimental frequencies and five associated $\delta f$) and $M=10$ adjusted constants ($\mu_{pd}/m_e$, $m_p/m_e$, $R_{\infty}$, $r_p$, $r_d$, and the five $\delta f^{\rm(th)}$). 

Results are displayed in Tab.~\ref{tab:Five transitions with LSA}.
We see that the uncertainties ``saturate" once the experimental frequency uncertainties are at the 10\,Hz level. 
There is no substantial reduction of the uncertainty of the reduced proton-deuteron mass ratio compared to the case of only three HD$^+$ transitions. But a strong (six-fold) reduction in the uncertainty of the proton-electron mass ratio compared to the case of only two H$_2^+$ transitions is obtained. The uncertainty of the deuteron charge radius is now similar to the CODATA 2018 value. We may compare Table~\ref{tab:Five transitions with LSA} with the corresponding Table ~\ref{tab:Five transitions} of the analytical model. The latter overestimates the projected reduction of uncertainties by a few orders, for the smallest assumed experimental uncertainties. This is due to the assumption of perfect correlations.

\begin{table}[tb!]
$\begin{array}{|c|c|c|c|c|c|c|c|c|c|c|}
\hline
\multicolumn{5}{|c|} {\text{Trans.}} & u_a,u_b,u_c,u_d,u_e & u_r (\lambda_{pd}) & u_r(m_p/m_e) & u_r (R_{\infty}) & u\left(r_p\right) & u\left(r_d\right) \\

\text{a} & \text{b} & \text{c} & \text{d} & \text{e} & \text{(kHz)} & \left(10^{-12}\right) & \left(10^{-12}\right) & \left(10^{-12}\right) & \text{(fm)} & \text{(fm)} \\

\hline
 3 & 4 & 5 & 1 & 2 &\{0.3,0.3,0.3,0.3,0.1\} & 3.8 & 8.9 & 43 & 0.068 & 0.0019 \\
\hline
 3 & 4 & 5 & 1 & 2 & \{0.1,0.1,0.1,0.1,0.03\} & 2.8 & 3.0 & 23 & 0.036 & 0.00099 \\
\hline
 3 & 4 & 5 & 1 & 2 & \{0.03,0.03,0.03,0.03,0.03\} & 2.7 & 1.3 & 19 & 0.030 & 0.00085 \\
\hline
 3 & 4 & 5 & 1 & 2 & \{0.03,0.03,0.03,0.01,0.01\} & 2.7 & 1.1 & 19 & 0.030 & 0.00082 \\
\hline
 3 & 4 & 5 & 1 & 2 & \{0.003,0.003,0.003,0.003,0.003\} & 2.7 & 0.68 & 19 & 0.030 & 0.00080 \\
 \hline
 3 & 4 & 5 & 1 & 2 & \{0.003,0.003,0.003,0.001,0.001\} & 2.7 & 0.67 & 19 & 0.030 & 0.00080 \\
\hline
 3 & 4 & 5 & 1 & 2 & \{0.001,0.001,0.001,0.001,0.001\} & 2.7 & 0.67 & 19 & 0.030 & 0.00080 \\
\hline
\hline
\multicolumn{6}{|c|}{\text{CODATA 2018}}& 46 & 60 & 1.9 & 0.0019 & 0.00074 \\
 \hline
\end{array}$
\caption{LSA determination of the five quantities: $\lambda_{pd}
=\mu_{pd}/m_e$, $m_p/m_e$, Rydberg constant, proton and deuteron chargee radii, by measuring five transitions, three in HD$^+$ (a, b, c) and two in H$_2^+$ (d, e). No other input data is included. Columns~1, 2, 3 indicate the chosen transitions of HD$^+$; columns~4, 5 refer to H$_2^+$. The transition labels are defined in previous tables. Column~6 gives the assumed uncertainties of the measured frequencies. Fractional or absolute uncertainties of the five adjusted constants 
are shown in columns~7-11.
}  \label{tab:Five transitions with LSA}
\end{table}

Alternatively, the HD$^+$ data can be combined with the H(1S-2S) and H-D(1S-2S) measurements to get more accurate determinations. In this case, the number of input data is $N=15$ (adding the three H/D experimental frequencies and two associated $\delta f$ with respect to the previous adjustment), whereas the number of adjusted constants is $M=12$ (adding the two H/D $\delta f^{\rm th}$). 

Results are displayed in Tab.~\ref{tab:Five transitions plus H-D with LSA}.
When the experimental uncertainties are assumed to be 1~Hz (data row 9 in the table), i.e. fractionally  $2\times10^{-14}$ to $2\times10^{-15}$, depending on the transition, the uncertainties plunge by impressive factors compared to CODATA 2018:

$u(\mu_{pd}/m_e)$ by a factor 600, 

$u(m_p/m_e)$ by a factor~1000, 

$u(R_\infty)$ by a factor~3.5,

$u(r_p)$ by a factor~11, 

$u(r_d)$ by a factor~14.

Compared to two scenarios discussed earlier, (i), (iii) in ref.\,\cite{Karr2016a},  

the reduction of the uncertainties $u(m_p/m_e)$, $u(\mu_{pd}/m_e)$, 
$u(R_\infty)$ $u(r_p)$, $u(r_d)$ is by 
factors~
72, 
110, 
6, 
37, 
140, 
respectively, where now experimental uncertainties are assumed to be two orders smaller, a different set of MHI transitions, and inclusion of different hydrogen data are considered.  

It should be noted that already for experimental uncertainties on the order of $(2-5)\times10^{-13}$ - one order lower than today - (data row 4 in the table) the uncertainties of the charge radii {and of the Rydberg constant} would be a factor 2 smaller than CODATA 2018 uncertainties, and the mass ratio uncertainties a factor $\simeq25$ smaller.

Finally, Table~\ref{tab:Five transitions plus H-D with LSA} also considers, in the last two data rows, a possible substantial reduction in QED theory uncertainty. This would mainly reduce the uncertainty of $r_d$.

\begin{table}[h!]
$\begin{array}{|c|c|c|c|c|c|c|c|c|c|c|}
\hline
\multicolumn{5}{|c|} {\text{Trans.}} & u_a,u_b,u_c,u_d,u_e & u_r (\lambda_{pd}) & u_r(\lambda_p) & u_r (R_{\infty}) & u\left(r_p\right) & u\left(r_d\right) \\

\text{a} & \text{b} & \text{c} & \text{d} & \text{e} & \text{(kHz)} & \left(10^{-12}\right) & \left(10^{-12}\right) & \left(10^{-12}\right) & \text{(fm)} & \text{(fm)} \\

\hline
2 & 3 & 4 & 1 & 2 & \{0.3,0.3,0.3,0.3,0.1\} & 29 & 6.4 & 15 & 0.016 & 0.0062 \\
\hline
2 & 3 & 4 & 1 & 2 & \{0.003,0.003,0.003,0.003,0.003\} & 2.7 & 0.73 & 1.4 & 0.0014 & 0.00053 \\
\hline
 3 & 4 & 5 & 1 & 2 & \{0.3,0.3,0.3,0.3,0.1\} & 3.1 & 5.1 & 2.7 & 0.0029 & 0.0011 \\
\hline
 3 & 4 & 5 & 1 & 2 & \{0.1,0.1,0.1,0.1,0.03\} & 1.4 & 2.4 & 1.0 & 0.00096 & 0.00037 \\
\hline
 3 & 4 & 5 & 1 & 2 & \{0.03,0.03,0.03,0.03,0.03\} & 0.47 & 1.1 & 0.78 & 0.00060 & 0.00023 \\
\hline
 3 & 4 & 5 & 1 & 2 & \{0.03,0.03,0.03,0.01,0.01\} & 0.47 & 0.76 & 0.61 & 0.00031 & 0.00011 \\
\hline
 3 & 4 & 5 & 1 & 2 & \{0.003,0.003,0.003,0.003,0.003\} & 0.088 & 0.12 & 0.57 & 0.00019 & 0.000055 \\
 \hline
 3 & 4 & 5 & 1 & 2 & \{0.003,0.003,0.003,0.001,0.001\} & 0.088 & 0.089 & 0.56 & 0.00019 & 0.000051 \\
\hline
 3 & 4 & 5 & 1 & 2 & \{0.001,0.001,0.001,0.001,0.001\} & 0.076 & 0.057 & 0.56 & 0.00019 & 0.000051 \\
 \hline
 \hline
3 & 4 & 5 & 1 & 2 &  
\{0.003, 0.003, 0.003, 0.003, 0.003\} & 0.080 & 0.11 & 0.48 & 0.00015 & 0.000025 \\
\hline
  3 & 4 & 5 & 1 & 2 & \{0.001,0.001,0.001,0.001,0.001\} & 0.066 & 0.052 & 0.48 & 0.00014 & 0.000013\\
\hline
\hline
\multicolumn{6}{|c|}{\text{CODATA 2018}}& 46 & 60 & 1.9 & 0.0019 & 0.00074 \\
\hline
\end{array}$
\caption{LSA: similar to Tab.~\ref{tab:Five transitions with LSA}, but with H(1S-2S) and H-D(1S-2S) measurements included as input data.
The last two scenarios are computed for QED theory uncertainties $1\times10^{-12}$, a factor 8 smaller than elsewhere in this work.  The main effect is a reduction of the uncertainty of $r_d$.
} \label{tab:Five transitions plus H-D with LSA}
\end{table}

\newpage
\subsubsection{Transitions with weak sensitivity to the mass ratios}
\label{sec:Weakly sensitive transitions}

Related to the fact that in any species there are transitions with positive and with negative sensitivity to the relevant mass ratios, there are also transitions with small, in a few cases very small sensitivity.
In H$_2^+$, transition~4 has sensitivity 29 times smaller than for transition~1. 
The latter may be viewed as a ``reference transition", exhibiting 
quasi-harmonic sensitivity, since its initial and final levels have small vibrational quantum numbers. Another transition, $(6,0)\rightarrow(13,2)$ has sensitivity $-6.504 \times 10^{-9}$\,at.u., approximately $1\times10^3$ times smaller. 
In HD$^+$, transition 6 has sensitivity approximately 110 times smaller than reference transition 2, according to the nonadiabatic calculation.

The existence of such transitions opens up an additional opportunity: determination of only the set of constants $(R_\infty, r_p, r_d)$,  ignoring the mass ratios.
In other words, the mass ratios are not adjusted in the LSA.
Keeping the present analysis simple, we may omit the CODATA 2018 values as input to the LSA, claiming that the current uncertainties of these constants are small enough that their significance in the results of the LSA would turn out to be small. This is certainly the case, if the most appropriate transitions are chosen. 
CODATA 2018 values of the mass ratios are still used to compute the theoretical transition frequencies that are input to the LSA.

We show in Tab.\,\ref{tab:Two weakly sensitive transitions plus H-D with LSA} a LSA example. In comparison with the result of Tab.\,\ref{tab:Five transitions with LSA} that relied on 5 transitions, here the obtained Rydberg constant uncertainty is similar,  while the radii's uncertainties are less than a factor 2 larger. All three uncertainties are substantially smaller than those of CODATA 2018. 

For the selected H$_2^+$ transition 4 the impact of the CODATA 2018 uncertainty of $m_p/m_e$ is $u_r(f_d)=2.7\times10^{-13}$ and thus is twice smaller than the fractional uncertainty of the adjusted Rydberg constant in {the third scenario of the table}. The above assumption is thus still acceptable. If necessary, one may select an H$_2^+$ transition with weaker sensitivity.

\begin{table}[h!]

\begin{centering}
$$\begin{array}{c|c|c|c|c|c}
 \multicolumn{2}{c|}{\text{Trans.}}  & u_a\text{, }u_d & u_r\left(R_{\infty }\right) & u(r_p) & u(r_d) \\
 \text{a} & \text{d} & \text{(kHz)} & \left(10^{-12}\right) & \text{(fm)} & \text{(fm)} \\
\hline
 6 & 4 & \{0.1,0.1\} & 2.5 & 0.0026 & 0.0010 \\
 6 & 4 & \{0.01,0.01\} & 0.63 & 0.00037 & 0.00013 \\
 6 & 4 & \{0.003,0.003\} & 0.58 & 0.00027 & 0.000090 \\
 6 & 4 & \{0.001,0.001\} & 0.58 & 0.00026 & 0.000085 \\
\hline
\hline
\multicolumn{3}{c|}{\text{CODATA 2018}} & 1.9 & 0.0019 & 0.00074 \\
\end{array}
$$
\par
\end{centering}

\caption{\label{tab:Two weakly sensitive transitions plus H-D with LSA}
LSA for the determination of Rydberg constant and charge radii only. Two MHI transitions with small sensitivity to the mass ratios are considered. The experimental inputs are:  one frequency of HD$^+$ ($a$, transition 6), one frequency of H$_2^+$ ($d$, transition 4), H(1S-2S) and H-D(1S-2S) isotope shift measurements. No input from CODATA 2018 is used in the LSA.
The uncertainties of the adjusted constants have saturated when the experimental uncertainties are reduced to 0.003\,kHz.
 } 

\end{table}

\newpage
\subsubsection{Ratios of frequencies}
\label{sec:Frequency ratios}

When we consider different transitions in the same species or in different species, we find pairs that have similar sensitivity to the respective mass ratios. Furthermore, all transition frequencies are proportional to the Rydberg constant. Finally, the QED uncertainties are correlated. Then, we may construct ratios of frequencies in which these constants or contributions are fully or partially suppressed. Frequency ratios are conceptually simple and therefore they are attractive for illustration purposes. In ref.\,\cite{Alighanbari2023} ratios of (experimentally available) HD$^+$ frequencies were considered. Because appropriate ratios  suppress the influence of QED uncertainties, it was argued that they can be used as figures of merit for a test of quantum mechanics.

Here, we consider ratios of HD$^+$ and H$_2^+$ frequencies.
By proper choice of the transition pair, such a ratio may remove the sensitivity to $m_p/m_e$. Indeed this sensitivity is present also in HD$^+$, where it is intertwined with the sensitivity to $m_d/m_p$. However, this latter ratio has been very precisely measured \cite{Fink2021} and its uncertainty is therefore of less concern.

By inspection, we found the HD$^+$ transitions 11: $(v=0,L=0)\rightarrow(4,1)$ and 12: $(v=1,L=0)\rightarrow(3,1)$ to have fractional sensitivities $\hat{s}_{m_p/m_e}=(f_{\rm nr})^{-1}\partial f_{\rm nr}/\partial(m_p/m_e)$ very similar to that of reference transition 1 in H$_2^+$ ($\hat{s}_{m_p/m_e}=-2.3948\times10^{-4}$). The HD$^+$ sensitivities are 
$\hat{s}_{m_p/m_e}=(-2.3791,-2.3907)\times10^{-4}$,
respectively.
The first HD$^+$ transition is particularly easily accessible and has already been utilized \cite{Schneider2010}.

The analysis of the fractional deviation of the HD$^+$(11) to H$_2^+$(1) frequency ratio,
\begin{equation}
    {[f_{11}({\rm HD}^+)/f_1({\rm H}_2^+)]_{\rm theo}\over
    [f_{11}({\rm HD}^+)/f_1({\rm H}_2^+)]_{\rm exp}}-1
    \label{eq:ratio}
\end{equation}
shows that the dominant  uncertainty contributions {of the theoretical ratio} are $
(1.2, 0.78, 0.33)\times10^{-12}$, originating from the uncertainties of $r_d$ (CODATA 2018), $m_d/m_p$ \cite{Fink2021}, and QED theory, respectively. (The contribution of $u(r_{p,2018})$ is smaller still.) The experimental frequency uncertainties are negligible in comparison, once smaller than 0.03\,kHz.

Thus, {in the context of today's knowledge of fundamental constants,} this frequency ratio directly probes the value of the deuteron charge radius. In contrast, in the ratio $f_5/f_1$ of two HD$^+$ frequencies discussed in ref.\,\cite{Alighanbari2023}, the dominant non-experiment uncertainties originate from QED theory and $\mu_{pd}/m_e$.

{If the improvements in the uncertainty of $r_d$ and of the mass ratios projected in the  previous sections are realized at a moderate level, a different interpretation may result. For improvements by factors of 7, the uncertainty of the theoretical ratio eq.\,(\ref{eq:ratio}),  $\simeq3\times10^{-13}$, would be dominated by today's QED uncertainty. Then, the comparison of experimental and calculated ratios would imply a test of quantum physics at this noteworthy level.}

\newpage
\subsubsection{Three species}
\label{sec:Three species}
We briefly discuss the scenario of three species that form a ``closed" triad, {i.e. having only two distinct nuclei}. We choose the non-radioactive triad H$_2^+$, D$_2^+$ and HD$^+$. We consider only transitions between low-lying vibrational levels, for which the QED contributions are more easily computable.

As can be seen from Tab.\,\ref{tab:LSA with 3 species}, data row 2, using one transition per species, already for the current theory uncertainty and assuming a {20-fold} improvement of experimental uncertainty compared to today, $m_p/m_e$, Rydberg constant and the proton and deuteron charge radii would be obtainable with competitive uncertainties, without using muonic hydrogen data as input data. 

Moreover, a putative improvement in QED theory uncertainty by a factor 8 would result in levels substantially below CODATA 2018 for all five fundamental constants. In the last scenario of the table, the fractional uncertainty $u_r(m_d/m_e)\simeq0.7\times10^{-12}$. 

\begin{table}[h!]

\begin{centering}
$$\begin{array}{c|c|c|c|c|c|c|c|c}
\multicolumn{3}{c|}{\text{Trans.}} & u_a\text{, }u_d\text{, }u_f & u_r\left(\mu _{\text{pd}}/m_e\right) & u_r\left(m_p\right/m_e) & u_r\left(R_{\infty }\right) & u(r_p) & u(r_d) \\
 \text{a} & \text{d} & \text{f} & \text{(kHz)} & \left(10^{-12}\right) & \left(10^{-12}\right) & \left(10^{-12}\right) & \text{(fm)} & \text{(fm)} \\
\hline
 3 & 1 & 5 & \{0.6,0.2,0.2\} & 19 & 19 & 5.9 & 0.0063 & 0.0025 \\
 3 & 1 & 5 & \{0.03,0.03,0.03\} & 18 & 19 & 1.5 & 0.0015 & 0.00060 \\
 3 & 1 & 5 & \{0.003,0.003,0.003\} & 18 & 19 & 1.5 & 0.0014 & 0.00056 \\
\hline
\hline
 3 & 1 & 5 & \{0.03,0.03,0.03\} & 2.6 & 2.7 & 0.79 & 0.00055 & 0.00022 \\
 3 & 1 & 5 & \{0.003,0.003,0.003\} & 2.6 & 2.7 & 0.63 & 0.00020 & 0.000085 \\
\hline
\hline
\multicolumn{4}{c|}{\text{CODATA 2018}}& 46 & 60 & 1.9 & 0.0019 & 0.00074 \\
\end{array}
$$

\end{centering}

\caption{
LSA for the determination of five fundamental constants from three MHI species and H/D data. The experimental inputs are:  one frequency of HD$^+$ (a: transition 3), one frequency of H$_2^+$ (d: transition 1), one frequency of D$_2^+$ (f: transition 5, $(1,0)\rightarrow(3,2)$), H(1S-2S) and H-D(1S-2S) measurements. 
No input from CODATA 2018 is used in the LSA.
The last two scenarios are computed for QED theory uncertainties $1\times10^{-12}$, a factor 8 smaller than elsewhere in this work. For both levels of theory uncertainty, the uncertainties of the adjusted constants saturate when the experimental uncertainties reach 0.003\,kHz. A correlation coefficient of 0.99 between transitions a-f and d-f has been assumed.
 \label{tab:LSA with 3 species}
}
\end{table}

\newpage
\section{Discussion and Conclusion}

Here, we derived two main results. First, it is in principle possible
to determine mass ratios 
vastly more accurately
than known today (CODATA 2018). This could be accomplished by, for example, measuring five MHI transitions with uncertainty at the 1\,Hz level. The set of transitions should include transitions between highly excited vibrational levels.

Second, also the Rydberg constant and charge radii can in principle be determined more accurately  than known today, provided that future MHI spectroscopy data is combined with already available H and D spectroscopy data.
Data on the charge radii from muonic
hydrogen spectroscopy is not required. The radii values are adjusted
to the MHI, H/D data.

In order to arrive at the above conclusions, we performed two different analyses.

The first is analytical model that was kept simple in order to highlight that one can take advantage of the correlated theory uncertainty for different transitions. The number of experimental transition data considered was kept small and equal to the number of unknown parameters to be determined in a particular scenario. 
We emphasized that in order to obtain accurate mass ratios, precise a priori knowledge
of the charge radii (from muonic hydrogen spectroscopy) is not essential. 
In this model, the values of the charge radii are fitted
to the data, always in conjunction with the QED corrections,
in form of the quantity $\Delta_{\rm nuc,QED}$. 

If, on the other
hand, one would take into account the muonic hydrogen spectroscopy charge radii, the QED corrections $\Delta_{\rm QED}$ and
the higher-order nuclear size corrections for the proton and the deuteron
$\Delta{}_{\rm nuc,h.o.}$ could be 
obtained separately, from eqs.\,(\ref{eq:Delta_nuc,QED for H2+},\ref{eq:Delta_nuc,QED(HD+)}). 

Within the analytical model we also showed that if more than one MHI species is measured, it
is in principle possible to obtain the differences of squared charge
radii of proton, deuteron and triton, with uncertainties comparable
with or smaller than CODATA 2018 - using data from MHI spectroscopy only.

The second analysis was a LSA. 
It has the advantage of being more flexible and powerful, for two reasons: first, it allows to take into account partial correlations of the theory uncertainties. Second, one can include fundamental constants results and/or data and theory from other systems. 
Such inclusion is extremely favourable, as is evident from the comparison of Tab.\,\ref{tab:Five transitions with LSA} and \ref{tab:Five transitions plus H-D with LSA}. It is then that an impressive improvement of accuracy of the 5 fundamental constants becomes possible in principle. If the projected fractional uncertainty $6\times10^{-14}$ for the proton-electron mass ratio would be achieved, this would be the most accurately determined fundamental constant, topping the electron g factor.

We highlight that one LSA scenario (Tab.\,\ref{tab:LSA with 3 species}) consists of just one highly accurate  measurement (10 Hz uncertainty level) on each member of the H$_2^+$ - HD$^+$ - D$_2^+$ triad, in combination with H/D data. This furnishes {uncertainties competitive with CODATA 2018 for} the 5 fundamental constants. Notably, the three vibrational transitions do not need to involve large vibrational quantum numbers $v$. {Thus, the computation of the theoretical frequencies {with the assumed uncertainty $8\times10^{-12}$} will be possible using {the already available QED theory techniques}.} 

In the presented LSAs, we only considered scenarios involving three of the six MHI,  HD$^+$, H$^+_2$, D$_2^+$, but obviously the treatment could  be extended to MHI that contain the triton. We expect that its properties can in principle also be determined with similar uncertainties as for the proton and the deuteron in the considered scenario.

Many scenarios relied on vibrational transitions between levels of large $v$. We caution that it will be challenging to perform the ab initio computation of the corresponding transition frequencies at an uncertainty level of $8\times10^{-12}$. It is especially the Bethe logarithm that is challenging to compute with sufficient accuracy~\citep{Korobov2012}. 

Even if the measurement scenarios we have considered allow to (partially) circumvent limitations associated with QED theory uncertainties, it remains beneficial to improve the theory further, as shown for example by the last two lines of Table~\ref{tab:LSA with 3 species}. This could be achieved through computation of higher-order corrections to the one-loop self-energy and two-loop corrections that are currently the largest sources of theoretical uncertainty~\cite{Korobov2021}, but also by recomputing in a three-body approach some corrections previously calculated in the adiabatic approximation, which would increase the correlations between uncertainties of different transition frequencies (see the discussion in Sec.~\ref{sec:theory uncertainty}).

In order to achieve the mentioned impressive uncertainties, we have considered experimental frequency uncertainties (systematic
and statistical combined) as small as 1~Hz, corresponding to fractional
frequency uncertainties in the $10^{-15}$ range. Such levels appear
achievable, as our earlier analyses have shown \citep{Schiller2014,Karr2014}.

Concerning the experimental feasibility of measuring a ``hot-band''
transition frequency, we point out that in HD\textsuperscript{+}
a rovibrational level with $v=9$ has a lifetime on the order of 10~ms.
This is long enough for allowing preparation of a MHI in this level
using e.g. Rabi flopping. The spectroscopic excitation should follow
within a time interval of order millisecond. In the homonuclear MHI,
the lifetime of all excited vibrational levels is of the order of
days (see ref.~\citep{Fink2021} for observations) and so the spectroscopy
can take place after a longer wait time, and with slower rate. 

In
${\rm H}_{2}^{+}$, the spectroscopy can be performed on electric quadrupole (E2) transitions 
\citep{Korobov2018a}. Recently, a vibrational transition has been observed \citep{Schenkel2023}, demonstrating feasibility. Transitions to be addressed in future work would likely be those with a small difference $v'-v$, in order to  achieve sufficiently high Rabi frequency with available laser sources.
Among transitions having $v'-v=2$, $(12,0)\rightarrow$ $(14,2)$
at 46~THz is one with positive mass ratio sensitivity, while $(9,0)\rightarrow$
$(11,2)$ at 68~THz has suppressed mass ratio sensitivity.
E2 transitions are also suitable for the vibrational spectroscopy of D$_2^+$, and have been discussed theoretically in detail \cite{Danev2020}.

In this work we also emphasized that powerful tests of consistency of experimental values obtained from different experiments 
may be performed: e.g.\,$r_p^2$ and $r_d^2$ obtained from H$_2^+$ and HD$^+$ must be consistent with the values obtained from the triad D$_2^+$, HT$^+$ and DT$^+$. Such tests could be very important in order to uncover overlooked systematic shifts and enhance confidence in the results.

The proposed approach leads to more accurate mass ratios via comparison between experiment and theory prediction. The values of these constants are  functions of the forces assumed to act between the particles contained in the MHI. Consequently, the approach
may also lead to more sensitive searches for beyond-Standard-Model (BSM)
forces between the particles, and more accurate
tests of their wave properties, topics that have been explored in
recent studies \citep{Germann2021,Alighanbari2020,Delaunay2023,Alighanbari2023}.
BSM signatures could appear in values of the obtained Rydberg
constant, mass ratios or squared radii differences, that do not agree
with measurements on the electronic and muonic hydrogen isotopes and
direct mass measurements. 
One simple example of a BSM physics test is the comparison of the charge radii from Tab.\,\ref{tab:Five transitions plus H-D with LSA}, obtained from MHI and electronic H/D, with those obtained from muonic H/D. The latter values are derived assuming conventional QED for the muon-radiation field interaction and for the muon-nucleus interaction. Any discrepancy between the two sets of results could hint at BSM forces. 

A BSM electron-nucleus  interaction that depends on the nuclear composition could be probed using data obtained
only from MHI
(as a violation of relationship eq.\,(3), SM)
or as inconsistent mass ratio values obtained from different MHI species. 
Such effects could also be tested in a LSA where the energy shifts induced by hypothetical BSM interactions are included, 
{and parameters describing these interactions are adjusted~\citep{Alighanbari2023,Delaunay2023}}.

\begin{acknowledgments}
We thank V.~I.~Korobov for putting at our disposal his codes for
the computation of the nonrelativistic energies and expectation values.
S\,S. thanks M.\,Schenkel, I.\,Kortunov and C.\,Wellers for discussions, and S.\,Alighanbari for discussions and comments on the text.
This work has received funding from the European Research Council
(ERC) under the European Union\textquoteright s Horizon 2020 research
and innovation programme (grant agreement No.~786306, ``PREMOL''
(S.S.))

\textbf{\smallskip{}
}
\end{acknowledgments}

\textbf{\smallskip{}
}


\vfill
\clearpage


\begin{center}
{\Large \textbf{Supplemental Material}}
\end{center}

\setcounter{equation}{0}  \renewcommand{\theequation}{SM\arabic{equation}}

\setcounter{table}{0}  \renewcommand{\thetable}{SM\arabic{table}}

\setcounter{figure}{0}  \renewcommand{\thefigure}{SM\arabic{figure}}

\vskip .1in


\section*{Analytical model: other scenarios}

In this Supplemental Material, we investigate the achievable accuracy of fundamental constants determinations in several measurement scenarios using the analytical model described in Sec.~II
of the main paper. The first three scenarios discussed below are also addressed in the main paper (see Sec.\,III.1, III.2
III.6)
using a more rigorous approach relying on least-squares adjustments.

\subsection{Two transitions in one species}

We first consider the case, when only two transitions have been measured,
$f_{a}^{\rm (expt)}$ and $f_{b}^{\rm (expt)}$. They have respective experimental
uncertainties $u_{a}=u(f_{a}^{\rm (expt)})$ and $u_{b}=u(f_{b}^{\rm (expt)})$,
which we assume to be uncorrelated. The uncertainty of the Rydberg
constant is taken as given, with value $u(\Delta_{h,2018})$ $=u_r(R_{\infty,2018})$. Solving
the equations
\[
f_{a}^{\rm (expt)}=f_{a}^{\rm (theor)},f_{b}^{\rm (expt)}=f_{b}^{\rm (theor)}\,,
\]
one immediately obtains $\Delta_{m,\mu}$, and it has an uncertainty
\begin{equation*}
u(\Delta_{m,\mu})^{2}  =(s_{b}V_{a}-s_{a}V_{b})^{-2}\times\Biggl(V_{b}^{2}\biggl(\frac{u_{a}}{2c\,R_{\infty}}\biggr)^{2}+V_{a}^{2}\biggl(\frac{u_{b}\,}{2c\,R_{\infty}}\biggr)^{2}+
\end{equation*}
 \begin{equation}
     \phantom{=+}\left((f_{\rm nr,a.u.})_{a}V_{b}\,-(f_{\rm nr,a.u.})_{b}V_{a}\right)^{2}u(\Delta_{h})^{2}\Biggr)\,\,,
     \label{eq:uncertainty of mu over me for 2 transitions}
\end{equation}
with the notations $V_{j}=(\langle V_{\delta,12}\rangle_{v',N'}-\langle V_{\delta,12}\rangle_{v,N})_{j}$
, $s_{j}=\partial(f_{\rm nr,a.u.})_{j}/\partial(\mu_{12}/m_{e})$ for
the two transitions $j=a,b$. The analogous expression for $\Delta_{\rm nuc,QED}$
will be omitted for brevity.

\begin{figure}[t]
\begin{centering}
\includegraphics[width=1\columnwidth]{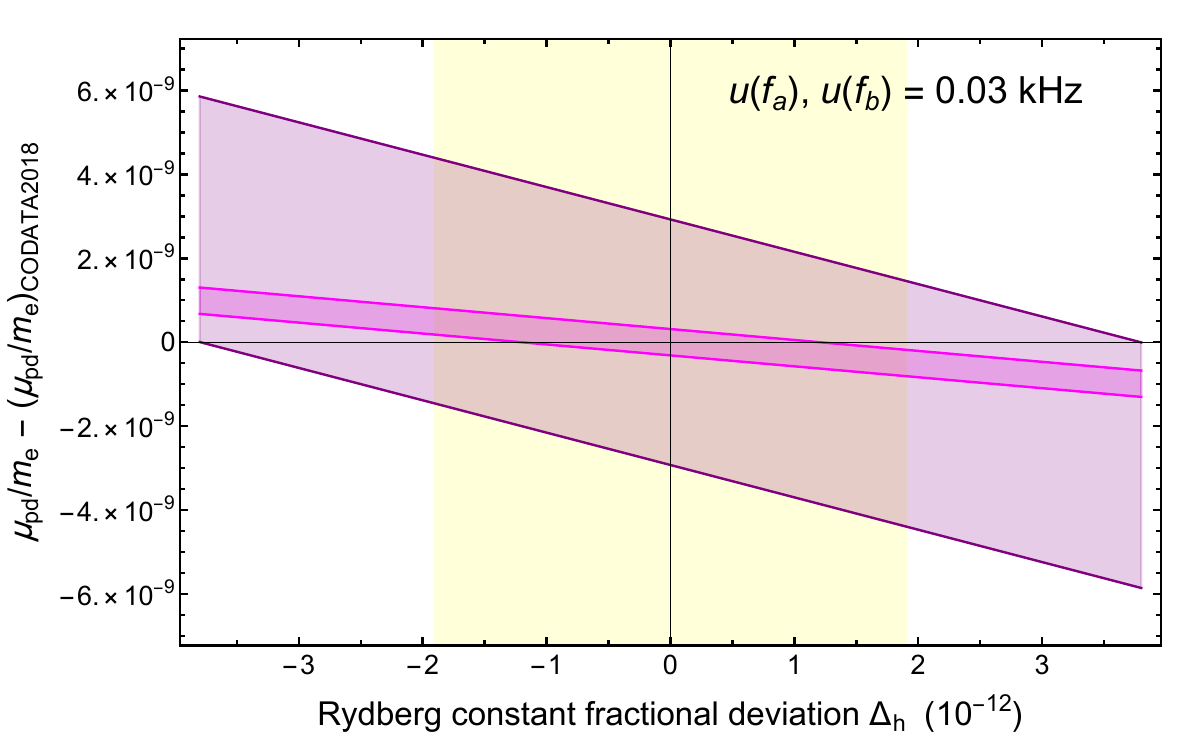}
\par\end{centering}
\caption{\label{fig:Determination-of-the mass ratio from 2 measurements}
Proposed
determination of the mass ratio $\mu_{pd}/m_{e}$ from measurements
of a frequency pair $(f_{a},f_{b})$ in HD\protect\textsuperscript{+}.
Two cases are shown, in purple and magenta. Purple: transitions $a{\rm:}\, (v=0,N=0)\rightarrow (v'=5,N'=1)$ and $b{\rm: }\,(0,3)\rightarrow(9,3)$.
Magenta: transitions $a{\rm:}\,(0,0)\rightarrow(5,1)$ and $b{\rm:}\, (9,1)\rightarrow(18,0)$. 
The widths of the magenta
and purple bands are due to the assumed experimental frequency uncertainties
$u(f_{a})$, $u(f_{b})$. 
The yellow band indicates today's (CODATA\,2018) uncertainty
of the Rydberg constant. The slope and width of a band together with
the width of the yellow band determines the uncertainty of $\mu_{pd}/m_{e}$. The magenta case is more favourable. }
\end{figure}

Table~\ref{tab:Two transitions with LSA} shows numerical examples of the
uncertainties achievable for different two-transition scenarios in
HD$^{+}$. One main result is that it is favourable to have a positive-sensitivity
transition result available (transition~5). Compared
to the case where no such result is available, one obtains the same
uncertainty $u(\Delta_{m,\mu})$ with a tenfold worse experimental
frequency uncertainty. Alternatively, for the same experimental frequency
uncertainty, in case of availability one obtains a five times smaller
uncertainty $u(\Delta_{m,\mu})$. As eq.~(\ref{eq:uncertainty of mu over me for 2 transitions})
indicates, the attainable uncertainty for the reduced mass ratio $\mu_{12}/m_{e}$
is limited by the uncertainty $u(\Delta_{h,2018})$ of
the Rydberg constant, to a value ${\rm Min}(u(\Delta_{m,\mu}))\approx6\times10^{-10}$.
This is a factor 100 smaller than the CODATA 2018 uncertainty and
a factor $\approx40$ smaller than the uncertainty reported to date from MHI spectroscopy, $u([\Delta_{m,\mu}]_{{\rm expt},{\rm HD}^{+}})\simeq2.5\times10^{-8}$.
Achieving the limit ${\rm Min}(u(\Delta_{m,\mu}))$ requires the experimental
uncertainties $u_{a},\,u_{b}$ to be approximately 0.03~kHz. Note
that this level is much more stringent than the transition frequency
uncertainty related to the Rydberg constant uncertainty, $u(\Delta_{h,2018})\times f_{a,b}\simeq 0.4\,{\rm kHz}$.
Figure~\ref{fig:Determination-of-the mass ratio from 2 measurements} displays
the advantage of using both a negative-sensitivity transition and a positive-sensitivity
transition (magenta band).

The last line in Table~\ref{tab:Two transitions with LSA} reports the result
for the experimentally measured transitions 3 and 4 when their actually
achieved experimental uncertainties are assumed. The computed uncertainty
$u(\Delta_{m,\mu})\simeq5.5\times10^{-8}$ is higher than the already
achieved uncertainty $u([\Delta_{m,\mu}]_{{\rm expt, HD}^+})$. This is so because
in the table, $\Delta_{\rm nuc,QED}$ was adjusted, while in the reported
works, it was not; $\Delta_{\rm nuc}$(HD$^+$) and $\Delta_{\rm QED}$ were set  to be zero with uncertainties as
quoted above.

\begin{table}[h!]

$$\begin{array}{|c|c|c|c|c|c|c|c|c|}
\hline
\multicolumn{2}{|c|}{} & \left|s_bV_a \!-\!s_aV_b\right|^{-1} & u_a,u_b & \multicolumn{4}{c|}{\text{analytical model}} & \text{LSA}\\ \cline{5-9}
 \multicolumn{2}{|c|}{} 
 & 
 & 
 & \multicolumn{2}{c|}{\text{contrib. to }u\left(\Delta _{m,\lambda }\right)\text{ from ...}} & u\left(\Delta _{m,\lambda }\right) & 
 u(\Delta_{{\rm HD}^+})
 & u\left(\Delta _{m,\lambda }\right)
 \\ 
 \text{a} & \text{b}& (10^6) & \text{(kHz)} & \text{}u_{{a}},u_{{b}}\left(10^{-10}\right) & \text{}u\left(\Delta _{h,2018}\right)\left(10^{-10}\right) & \left(10^{-10}\right) & 
 \text{} & {\left(10^{-10}\right)} \\
\hline
2 & 3 & 75 & 0.3 & 1500 & 23 & 1500 & 2.3 &233\\
\hline
 2 & 3 & 75 & 0.03 & 150 & 23 & 150 & 0.26 &180\\
\hline
 3 & 4 & 8.3 & 0.3 & 290 & 15 & 290 & 0.43 &202\\
\hline
 3 & 4 & 8.3 & 0.03 & 29 & 15 & 33 & 0.12 &136\\
\hline
 3 & 5 & 1.4 & 0.3 & 31 & 4.9 & 32 & 0.11 &43\\
\hline
 3 & 5 & 1.4 & 0.03 & 3.1 & 4.9 & 5.9 & 0.094 &29\\
\hline
 3 & 6 & 2.2 & 0.3 & 53 & 7.4 & 54 & 0.12 &91\\
\hline
 3 & 8 & 3.0 & 0.3 & 80 & 9.4 & 80 & 0.14 &125\\
\hline
 3 & 4 & 8.3 & \{0.6,0.46\} & 550 & 15 & 550 & 0.78&240 \\
 \hline
 \hline
\multicolumn{6}{|c|}{\text{CODATA 2018}}&  560 & & 560  \\
\hline
\end{array}
$$

\caption{
Examples of mass ratio determination by measuring a pair of transitions a and b in HD$^+$. In the analytical model no other data is used. The absolute uncertainty of $\lambda=\mu_{pd}/m_e$ 
is given in column 7 as $u(\Delta_{m,\lambda})$, and column 8 is the uncertainty of the fitted nuclear plus QED correction parameter. 
In the 
LSA procedure (last column), non-perfect theory uncertainty correlations are taken into account and, in addition to the two HD$^+$ data, also the CODATA 2018 values of $R_\infty$, $r_p$, $r_d$, see main paper, sec.\,III.1.
The last data row considers the scenario of two already performed experiments. We used the abbreviation 
$\Delta_{{\rm HD}^+}=\Delta _{\text{nuc},\text{QED}}(\text{HD}^+)$.}
\label{tab:Two transitions with LSA}
\end{table}

\newpage
\subsection{Three transitions in one species}

With three or more transitions available, besides $\Delta_{m,\mu}$
and $\Delta_{{\rm \rm nuc,QED}}$, also the Rydberg constant can be determined.
Explicit expressions for $u(\Delta_{m,\mu})$, $u(\Delta_{{\rm nuc,QED}})$ and
$u(\Delta_{h})$ can easily be derived, but will not be displayed
here. Table~\ref{tab:Three transitions} shows numerical results
for various combinations of transitions. For given experimental frequency
uncertainties $u_{a}$, $u_{b}$ and $u_{c}$, those combinations containing
the positive-mass-sensitivity transition 5 yield reduced uncertainties.
We find that the three uncertainties $u(\Delta_{m,\mu})$, $u(\Delta_{{\rm nuc,QED}})$, and
$u(\Delta_{h})$ drop continuously with decreasing $u_{a}$, $u_{b}$, and $u_{c}$.
In particular, the uncertainty of $\Delta_{m,\mu}$ drops below the
minimum achievable for the case of only two measured transitions.
For example, $u(\Delta_{m,\mu})\simeq4\times10^{-11}$, one order
lower than for the case of two transitions only, is obtained if the
experimental frequency uncertainties are 1~Hz. For such experimental
accuracy, the uncertainty of the Rydberg constant is determined with
ten-fold lower uncertainty than CODATA 2018's uncertainty.

We point out that the QED contribution parameter $\Delta_{{\rm nuc,QED}}$
is then determined with uncertainty 0.01, i.e. approximately 30 times
smaller that today's estimate. For an individual transition this corresponds
to a theoretical uncertainty of e.g.~$\simeq60$ Hz for transition~3,
a much larger value than the assumed experimental uncertainty. The
reason for the different magnitude is the correlation between the
QED deviation function $\langle V_{\delta,12}\rangle_{v',N'}-\langle V_{\delta,12}\rangle_{v,N}$
and the transition frequency $f(v,N\rightarrow v',N')$.

\begin{table}[t]
$\begin{array}{|c|c|c|c|c|c|c|c|c|c|}
\hline
\multicolumn{3}{|c|} {\text{Transitions}} & u_a,u_b,u_c & \multicolumn{3}{c|}{\text{contrib. to }u\left(\Delta _{m,\lambda }\right)\text{ from ...}}& u\left(\Delta _{m,\lambda }\right) & u\left(\Delta _h\right) & u\left(\Delta _{\text{nuc},\text{QED}}\right) \\
\text{a} & \text{b} & \text{c} & \text{(kHz)} & \text{}u_{{a}}\left(10^{-10}\right) & \text{}u_{{b}}\left(10^{-10}\right) & \text{}u_{{c}}\left(10^{-10}\right) & \left(10^{-10}\right) & \left(10^{-12}\right) & \text{} \\
\hline
 2 & 3 & 4 & \{0.15,0.6,0.46\} & 1300 & 700 & 2600 & 3000 & 330 & 20 \\
\hline
 3 & 4 & 5 & \{0.3,0.3,0.3\} & 98 & 39 & 81 & 130 & 55 & 2.7 \\
\hline
 3 & 4 & 5 & \{0.03,0.03,0.03\} & 9.8 & 3.9 & 8.1 & 13 & 5.5 & 0.27 \\
\hline
 3 & 4 & 5 & \{0.01,0.01,0.01\} & 3.3 & 1.3 & 2.7 & 4.4 & 1.8 & 0.091 \\
\hline
 3 & 4 & 5 & \{0.003,0.003,0.003\} & 0.98 & 0.39 & 0.81 & 1.3 & 0.55 & 0.027 \\
\hline
 3 & 4 & 5 & \{0.001,0.001,0.001\} & 0.33 & 0.13 & 0.27 & 0.44 & 0.18 & 0.0091 \\
\hline
 3 & 5 & 6 & \{0.003,0.003,0.003\} & 0.15 & 0.85 & 0.79 & 1.2 & 0.41 & 0.020 \\
\hline
 3 & 5 & 8 & \{0.003,0.003,0.003\} & 0.24 & 0.64 & 0.55 & 0.88 & 0.31 & 0.016 \\
\hline
 3 & 6 & 7 & \{0.003,0.003,0.003\} & 0.27 & 10 & 5.5 & 12 & 3.0 & 0.16 \\
\hline
 3 & 6 & 8 & \{0.003,0.003,0.003\} & 0.47 & 2.1 & 1.9 & 2.9 & 0.70 & 0.037 \\
\hline
\hline
\multicolumn{7}{|c|}{\text{CODATA 2018}}&  560 & 1.9 &  \\
\hline
\end{array}
$

\caption{\label{tab:Three transitions}
Analytical model: examples of the determination of both
the reduced mass $\lambda=\mu_{pd}/m_e$ and the Rydberg constant by measuring three transitions
in HD\protect\textsuperscript{+}. In column 8, $u(\Delta_{m,\lambda})$ is the absolute uncertainty of $\lambda$. In column~9, $u\left(\Delta _h\right)=$ $u_r(R_\infty)$ is the fractional uncertainty of the fitted Rydberg constant. The first case in the table (data row 1) considers
the three rovibrational transitions measured to date. In this
case, because the Rydberg constant is among the fitted constants,
no competitive uncertainty is obtained for the fitted mass ratio.}
\end{table}

\subsection{Three Species}

 Suppose precision spectroscopy is also performed on ${\rm D}_{2}^{+}$.
Data from it alone will provide the deuteron-electron mass ratio $m_{d}/m_{e}$,
and thus a consistency check is possible with the independent value
obtained from combined HD$^{+}$ and ${\rm H}_{2}^{+}$ measurements.
Furthermore, $\Delta_{\rm nuc,QED}({\rm D}_{2}^{+})$ is obtained. Since
\begin{align*}
\Delta_{\rm nuc,QED}({\rm D}_{2}^{+}) & =\alpha^{-5}(2\pi/3)a_{0}^{-2}\frac{1}{2}\times2\,\Delta(r_{d}^{2})+\Delta_{\rm QED}+2\Delta{}_{\rm nuc,h.o.}(d)\,,
\end{align*}
the comparison with 
eqs.\,(16,17) (main paper)
shows that
\begin{equation}
\Delta_{\rm nuc,QED}({\rm D}_{2}^{+})=2\Delta_{\rm nuc,QED}({\rm HD}^{+})-\Delta_{\rm nuc,QED}({\rm H}_{2}^{+})\,\,.\label{eq:Relationship between the Delta_nuc_QED}
\end{equation}
Thus, data from ${\rm D}_{2}^{+}$ does not provide new information
concerning the radii. One cannot simultaneously determine the unknown QED contribution and obtain the radii $r_{p}$ , $r_{d}$
\emph{individually}. Still, this expression provides a very important
consistency check with the result from combined ${\rm H}_{2}^{+}$,
HD$^{+}$ measurements. The three-species scenario is also analyzed in Sec.\,III.6
of the main paper, this time including data from atomic hydrogen spectroscopy. A different conclusion is then reached.

\subsection{Including one triton-containing species}

If e.g. HT\textsuperscript{+} is measured, apart from $\mu_{pt}/m_{e}$,
the fit also yields $\Delta_{\rm nuc,QED}({\rm HT}^{+})$, which is given
by
\begin{align*}
\Delta_{\rm nuc,QED}({\rm HT}^{+}) & =\alpha^{-5}(2\pi/3)a_{0}^{-2}\frac{1}{2}\times\Delta(r_{p}^{2}+r_{t}^{2})+\Delta_{\rm QED}+\Delta{}_{\rm nuc,h.o.}(t)\,.
\end{align*}
The recoil correction of this species is assumed to be treated in the same wave as described earlier. One can combine this with the H$_2^+$ results, 
eq.\,(16) in the main paper, to
obtain $\Delta(r_{t}^{2})-\Delta(r_{p}^{2})$ with minimum uncertainty
given by the uncertainty of $\Delta{}_{\rm nuc,h.o.}(t)$. 
The latter is dominated by the triton polarizability correction $E_{\rm pol}^{(5)}$, for which no calculation has been reported, although the electric dipole polarizability of the triton was calculated in~\citep{Stetcu2009}. One may adopt the estimate given in~\citep{Yerokhin2015}:
\begin{align*}
E_{\rm pol}^{(5)} \approx - \frac{E_{\rm fns}}{1000} \pm 100 \% \,,   
\end{align*}
where $E_{\rm fns}$ is the leading-order finite-nuclear-size correction. This yields
\begin{align*}
u(\delta f_{\rm nuc,h.o.}(t)) &= \frac{1}{1000} \times 2cR_{\infty} \frac{2\pi}{3} \left(\frac{r_t}{a_0}\right)^{2} \frac{1}{2}
\left(\langle V_{\delta,pt}\rangle_{v',N'}-\langle V_{\delta,pt}\rangle_{v,N}\right)\, , \nonumber \\
&= 7.6\,{\rm kHz}\times(\langle V_{\delta,pt}\rangle_{v',N'}-\langle V_{\delta,pt}\rangle_{v,N})\,;\nonumber\\    
u(\Delta_{\rm nuc,h.o.}(t)) &=\frac{1}{1000} \times  \alpha^{-5}\frac{2\pi}{3} \left(\frac{r_t}{a_0}\right)^{2} \frac{1}{2}\ .
\end{align*}
This implies that $\Delta(r_{t}^{2})-\Delta(r_{p}^{2})$ could then be deduced
with an uncertainty of about $r_t^2/1000\simeq0.003\,{\rm fm^{2}}$, 
which would already represent a considerable
improvement compared to today's experimental uncertainty.
The current triton radius value is $r_{t}^{{\rm (expt)}}=1.755(86)\,{\rm fm}$
from electron scattering experiments \citep{Amroun1994}. 
Its absolute uncertainty is substantially larger than that of $r_{p}$ and $r_{d}$,
eq.\,(3) in the main paper.
Calculation of the triton polarizability contribution would allow reaching even higher precision.

In nuclear physics, the triton is a nucleus of substantial interest.
Its \emph{point} charge radius $\delta r_{C}$, that is related to
the charge radius, can be computed using effective field theory \citep{Vanasse2017,Filin2023}
with competitive uncertainty. A recent calculation of $\delta r_{C}^{{\rm (theory)}}$
yields, when combined with the experimental $^{3}$He charge radius,
a preliminary value $r_{t}=1.773(9)\,{\rm fm}$ \citep{Filin2023}.
Precision data on H$_{2}^{+}$ and HT$^{+}$ could verify this prediction
and also future, much improved predictions. Thus, HT$^+$ spectroscopy could be an alternative to future laser spectroscopy of atomic tritium \cite{Schmidt2018}.

\subsection{The group of heteronuclear MHI}

If DT\textsuperscript{+} is also included to complete the triad of heteronuclear MHI, one will obtain  $\mu_{dt}/m_{e}$ and $\Delta_{\rm nuc,QED}({\rm DT}^{+})$, given by
\begin{align*}
\alpha^{5}\Delta_{\rm nuc,QED}({\rm DT}^{+}) & =(2\pi/3)a_{0}^{-2}\frac{1}{2}\times\Delta(r_{d}^{2}+r_{t}^{2})+\Delta_{\rm QED}+\Delta{}_{\rm nuc,h.o.}(d)+\Delta{}_{\rm nuc,h.o.}(t)\,.
\end{align*}
Combination with the HT\textsuperscript{+} result yields
\begin{align*}
\alpha^{5}\Delta_{\rm nuc,QED}({\rm DT}^{+}) - \alpha^{5}\Delta_{\rm nuc,QED}({\rm HT}^{+}) & =(2\pi/3)a_{0}^{-2}\frac{1}{2}\times\Delta(r_{d}^{2}-r_{p}^{2})+\Delta{}_{\rm nuc,h.o.}(d)\,.
\end{align*}
This is the same result as eq.\,(18)
 in the main paper,
obtained from H$_2^+$ and HD\textsuperscript{+}. Similarly, combining DT\textsuperscript{+} and HD\textsuperscript{+} results one obtains $\Delta(r_{t}^{2}-r_{d}^{2})$ (modulo the higher-order nuclear correction for $t$). This means that one can in principle obtain the information on the charge radii without performing measurements on the homonuclear ions. Such an approach may have some experimental advantages. Alternatively, one can verify experimentally consistency of  independent measurements performed on four different MHI, e.g. H$_2^+$, HD\textsuperscript{+}, HT\textsuperscript{+}, DT\textsuperscript{+}. 


\vfill\clearpage


\begin{center}
{\Large \textbf{Erratum}}
\end{center}
\vskip .1in

\setcounter{equation}{0}  \renewcommand{\theequation}{E\arabic{equation}}

\setcounter{table}{0}  \renewcommand{\thetable}{E\arabic{table}}

\setcounter{figure}{0}  \renewcommand{\thefigure}{E\arabic{figure}}

In our publication \cite{Schiller2024}, in the sensitivities of the H$_2^+$ transition frequencies to the proton charge radius $r_p$ ($\beta_{r_p}$, see Eq.~(19)), a factor 2 was unfortunately missing. As a consequence, the improvement factors of the determination of the fundamental constants are not as strong any more. 

However, we point out that in other scenarios, described in a forthcoming publication \cite{Karr2025}, we find that further improvements occur compared to the here presented revised numbers.

The original text requires several corrections. 
(No revision is required in the Supplemental Material.)

 In particular, this statement in the abstract does not hold anymore: ``allowing us in principle to reach uncertainties for the mass ratios approximately three orders smaller than reported by CODATA 2018. Improvements by a factor of 3.5 for the Rydberg constant, and 11 (14) for the proton (deuteron) charge radius, are also projected".

For the mass ratios, the improvements are factors 150 -- 300, while for Rydberg constant, proton and deuteron charge radius approximately the same uncertainties as for CODATA 2018 can in principle be reached, see the corrected Table VIII in Tab.\,E3.

 The last two sentences of the abstract should now read:
``We find these to be powerful approaches, allowing us in principle to reach uncertainties for the mass ratios smaller by two orders compared to CODATA 2018. For the Rydberg constant, proton and deuteron charge radii approximately the same uncertainties are reachable, without use of muonic H/D data."

\vskip .1in
The numbers in tables V, VII, VIII, IX, X are incorrect and need to be revised. The revised tables are given below.
The comments on the various tables also need to be modified
as follows.

\begin{itemize}
{\item Corrected Table V: the last sentence of Sec. III.A should read: \\
``For that case, the mass-ratio uncertainty, of order approximately $5.0 \times 10^{-12}$ fractionally, is 12 times smaller 
than the CODATA 2018 uncertainty.''}

{\item Corrected Table VII: the beginning of the second paragraph of Sec. III.C should read: \\ ``Results are displayed in Table VII. We see that the uncertainties ``saturate'' once the experimental frequency uncertainties are at the 30-Hz level. There is no substantial reduction of the uncertainty of the reduced proton-deuteron mass ratio compared to the case of only three HD$^+$ transitions; however, a twofold reduction in the uncertainty of the proton-electron mass ratio compared to the case of only two H$_2^+$ transitions is obtained.'' The rest of the paragraph (``We may compare Table VII with...'') is unchanged.}

{\item Corrected Table VIII: the end of Sec. III.C, starting from the fourth paragraph, should be modified to:\\
``Results are displayed in Table VIII. All the uncertainties are significantly reduced by the inclusion of hydrogen atom data. When the experimental uncertainties are assumed to be 3 Hz (data row~7 in the table), i.e., fractionally $6 \times 10^{-14}$ to $6 \times 10^{-15}$, depending on the transition, improvement factors compared to CODATA 2018 reach 60 for $\mu_{pd}/m_e$ and 120 for $m_p/m_e$. \\
Compared to a scenario discussed earlier that also involves 3 HD$^+$ and 2 H$_2^+$ transitions, (i) in Ref. [12], the reduction of the uncertainties $u(m_p/m_e)$, $u(m_d/m_p)$, $u(R_{\infty})$, $u(r_p)$, and $u(r_d)$ is by factors 18, 
67, 1.5, 1.2, and 2.8, respectively, where now experimental uncertainties are assumed to be two orders smaller, theoretical uncertainties a factor 2.5 larger, a different set of MHI transitions is considered, and hydrogen data are included. 

Finally, Table VIII also considers, in the last two data rows, a possible substantial reduction in QED theory uncertainty. This would reduce all the uncertainties further by factors of 2.5 to 3. Without input from muonic H/D data, proton and deuteron charge radii uncertainties are then close to CODATA 2018 levels.}

{\item Corrected Table IX: the last two paragraphs of Sec. III.D should be replaced with:\\
``We show in Table IX a LSA example. In comparison with the result of Table VII that relied on five transitions, here the obtained uncertainties for the Rydberg constant and radii are smaller by factors of 1.5 to 3.''}

{\item Corrected Table X: the last two paragraphs of Sec. III.F should be replaced with:\\
''As can be seen from Table X, data row 2, using one transition per species, already for the current theory uncertainty and assuming a 20-fold improvement of experimental uncertainty
compared to today, the mass ratios would be obtainable with competitive uncertainties. Moreover, a putative improvement in QED theory uncertainty by a factor 8 would result in levels substantially below CODATA 2018.}
\end{itemize}

Finally, some parts of the Conclusion (Sec.\,IV) need to be modified:
\begin{itemize}
{\item Second paragraph, first sentence:\\
``Second, also the Rydberg constant and charge radii can in principle be determined with competitive accuracy, provided that future MHI spectroscopy data are combined with already available H and D spectroscopy data.''}
{\item Fifth paragraph, last sentence:\\
`` If the projected fractional uncertainty $1.8 \times 10^{-13}$ for the proton-electron mass
ratio would be achieved, this would be one of the most accurately determined fundamental constant, only rivaled by the electron g-factor.''}
{\item Sixth paragraph:\\
``We highlight that one LSA scenario (Table X) consists of just one highly accurate measurement (3-Hz uncertainty level) on each member of the H$_2^+$-HD$^+$-D$_2^+$ triad, in combination with H and D data. Assuming reduction of QED theory uncertainty by a factor of 8, it furnishes improved determinations of the mass ratios. Notably, the three vibrational transitions do not need
to involve large vibrational quantum numbers $v$. Thus, the computation of the theoretical frequencies with the assumed uncertainty would not be hampered by numerical difficulties.''}

\end{itemize}

\renewcommand{\arraystretch}{0.6}
\begin{table}[h!]
    \centering
$\begin{array}{cccccccc}
\hline
  \multicolumn{2}{c}{\text{Transitions}}& u_a\text{, }u_b & u_r\left(m_p\right/m_e) & u_r\left(R_{\infty }\right) & \left.u(r_p\right) \\
 \text{a} & \text{b} & \text{(kHz)} & \left(10^{-12}\right) & \left(10^{-12}\right) & \text{(fm)} \\
\hline
 1 & 3 & 0.3,0.3 & 9.1 & 23 & 0.024 \\
 1 & 3 & 0.1,0.1 & 5.0 & 21 & 0.023 \\
 1 & 3 & 0.03,0.03 & 4.3 & 21 & 0.022 \\
 1 & 3 & 0.01,0.01 & 4.3 & 21 & 0.022 \\
 1 & 3 & 0.003,0.003 & 4.3 & 21 & 0.022 \\
 1 & 3 & 0.001,0.001 & 4.3 & 21 & 0.022 \\
\hline
 \multicolumn{3}{c}{\text{CODATA 2018}}& 60 & 1.9 & 0.0019 \\
\hline
\end{array}
$
    \caption{Corrected Table V. Input data are two transitions (a and b) in H$_2^+$ and H(1S-2S).}
    \label{tab:Corrected Table V}
\end{table}

\begin{table}[h!]
    \centering
\small
$\begin{array}{ccccccccccc}
\hline
  \multicolumn{5}{c}{\text{Transitions}}& u_a\text{, }u_b\text{, }u_c\text{, }u_d\text{, }u_e & u_r\left(\mu _{\text{pd}}\right/m_e) & u_r\left(m_p\right/m_e) & u_r\left(R_{\infty }\right) & \left.{u(}r_p\right) & \left.{u(}r_d\right) \\
 \text{a} & \text{b} & \text{c} & \text{d} & \text{e} & \text{(kHz)} & \left(10^{-12}\right) & \left(10^{-12}\right) & \left(10^{-12}\right) & \text{(fm)} & \text{(fm)} \\
\hline
 3 & 4 & 5 & 1 & 2 & 0.3,0.3,0.3,0.3,0.1 & 12 & 12 & 62 & 0.049 & 0.021 \\
 3 & 4 & 5 & 1 & 2 & 0.1,0.1,0.1,0.1,0.03 & 5.2 & 4.6 & 33 & 0.026 & 0.011 \\
 3 & 4 & 5 & 1 & 2 & 0.03,0.03,0.03,0.03,0.03 & 3.7 & 2.6 & 28 & 0.022 & 0.0091 \\
 3 & 4 & 5 & 1 & 2 & 0.03,0.03,0.03,0.01,0.01 & 3.7 & 2.5 & 28 & 0.022 & 0.0091 \\
 3 & 4 & 5 & 1 & 2 & 0.003,0.003,0.003,0.003,0.003 & 3.5 & 2.2 & 27 & 0.022 & 0.0089 \\
 3 & 4 & 5 & 1 & 2 & 0.003,0.003,0.003,0.001,0.001 & 3.5 & 2.2 & 27 & 0.022 & 0.0089 \\
 3 & 4 & 5 & 1 & 2 & 0.001,0.001,0.001,0.001,0.001 & 3.5 & 2.2 & 27 & 0.022 & 0.0089 \\
\hline
 \multicolumn{6}{c}{\text{CODATA 2018}}& 46 & 60 & 1.9 & 0.0019 & 0.00074 \\
\hline
\end{array}
$
    \caption{Corrected Table VII. The input data consists of five transitions, three in HD$^+$ (a, b, c) and two in H$_2^+$ (d and e).}
    \label{tab:Corrected Table VII}
\end{table}

\begin{table}[h!]
\begin{small}
    \centering
$\begin{array}{ccccccccccc}
\hline
  \multicolumn{5}{c}{\text{Transitions}}& u_a\text{, }u_b\text{, }u_c\text{, }u_d\text{, }u_e & u_r\left(\mu _{\text{pd}}\right/m_e) & u_r\left(m_p\right/m_e) & u_r\left(R_{\infty }\right) & \left.{u(}r_p\right) & \left.{u(}r_d\right) \\
 \text{a} & \text{b} & \text{c} & \text{d} & \text{e} & \text{(kHz)} & \left(10^{-12}\right) & \left(10^{-12}\right) & \left(10^{-12}\right) & \text{(fm)} & \text{(fm)} \\
\hline
 2 & 3 & 4 & 1 & 2 & 0.3,0.3,0.3,0.3,0.1 & 5.7 & 6.3 & 21 & 0.022 & 0.0087 \\
 2 & 3 & 4 & 1 & 2 & 0.003,0.003,0.003,0.003,0.003 & 0.80 & 0.71 & 11 & 0.012 & 0.0048 \\
 3 & 4 & 5 & 1 & 2 & 0.3,0.3,0.3,0.3,0.1 & 4.6 & 6.4 & 21 & 0.022 & 0.0088 \\
 3 & 4 & 5 & 1 & 2 & 0.1,0.1,0.1,0.1,0.03 & 2.3 & 2.7 & 15 & 0.016 & 0.0062 \\
 3 & 4 & 5 & 1 & 2 & 0.03,0.03,0.03,0.03,0.03 & 1.6 & 1.7 & 12 & 0.013 & 0.0052 \\
 3 & 4 & 5 & 1 & 2 & 0.03,0.03,0.03,0.01,0.01 & 1.0 & 1.0 & 8.2 & 0.0087 & 0.0034 \\
 3 & 4 & 5 & 1 & 2 & 0.003,0.003,0.003,0.003,0.003 & 0.74 & 0.49 & 5.8 & 0.0060 & 0.0024 \\
 3 & 4 & 5 & 1 & 2 & 0.003,0.003,0.003,0.001,0.001 & 0.72 & 0.45 & 5.6 & 0.0058 & 0.0024 \\
 3 & 4 & 5 & 1 & 2 & 0.001,0.001,0.001,0.001,0.001 & 0.72 & 0.45 & 5.6 & 0.0058 & 0.0023 \\
\hline
 3 & 4 & 5 & 1 & 2 & 0.003,0.003,0.003,0.003,0.003 & 0.29 & 0.22 & 2.2 & 0.0021 & 0.00086 \\
 3 & 4 & 5 & 1 & 2 & 0.001,0.001,0.001,0.001,0.001 & 0.28 & 0.18 & 2.2 & 0.0021 & 0.00085 \\
\hline
 \multicolumn{6}{c}{\text{CODATA 2018}}& 46 & 60 & 1.9 & 0.0019 & 0.00074 \\
\hline
\end{array}
$
    \caption{Corrected Table VIII. Three transitions in HD$^+$ (a, b, c), two in H$_2^+$ (d and e), H(1S-2S) and D(1S-2S) are included as input data.}
    \end{small}
    \label{tab:Corrected Table VIII}
\end{table}


\begin{table}[h!]
    \centering
$\begin{array}{cccccc}
\hline
  \multicolumn{2}{c}{\text{Transitions}}& u_a\text{, }u_d & u_r\left(R_{\infty }\right) & \left.{u(}r_p\right) & \left.{u(}r_d\right) \\
 \text{a} & \text{d} & \text{(kHz)} & \left(10^{-12}\right) & \text{(fm)} & \text{(fm)} \\
\hline
 6 & 4 & 0.1,0.1 & 28 & 0.030 & 0.012 \\
 6 & 4 & 0.01,0.01 & 12 & 0.013 & 0.0050 \\
 6 & 4 & 0.003,0.003 & 9.1 & 0.0096 & 0.0038 \\
 6 & 4 & 0.001,0.001 & 8.8 & 0.0093 & 0.0037 \\
\hline
 \multicolumn{3}{c}{\text{CODATA 2018}}& 1.9 & 0.0019 & 0.00074 \\
 \hline
\end{array}
$
    \caption{Corrected Table IX. One frequency
of HD$^+$ (a), one frequency of H$_2^+$ (d), H(1S-2S) and H-D(1S-2S) are included as input data.
}
    \label{tab:Corrected Table IX}
\end{table}

\begin{table}[h!]
    \centering
$\begin{array}{ccccccccc}
\hline
  \multicolumn{3}{c}{\text{Transitions}}& u_a\text{, }u_d\text{, }u_f & u_r\left(\mu _{\text{pd}}\right/m_e) & u_r\left(m_p\right/m_e) & u_r\left(R_{\infty }\right) & \left.{u(}r_p\right) & \left.{u(}r_d\right) \\
 \text{a} & \text{d} & \text{f} & \text{(kHz)} & \left(10^{-12}\right) & \left(10^{-12}\right) & \left(10^{-12}\right) & \text{(fm)} & \text{(fm)} \\
\hline
 3 & 1 & 5 & 0.6,0.2,0.2 & 290 & 300 & 280 & 0.30 & 0.12 \\
 3 & 1 & 5 & 0.03,0.03,0.03 & 79 & 80 & 68 & 0.073 & 0.029 \\
 3 & 1 & 5 & 0.003,0.003,0.003 & 76 & 77 & 64 & 0.069 & 0.027 \\
\hline
 3 & 1 & 5 & 0.03,0.03,0.03 & 26 & 26 & 25 & 0.026 & 0.010 \\
 3 & 1 & 5 & 0.003,0.003,0.003 & 11 & 11 & 9.1 & 0.0099 & 0.0039 \\
\hline
 \multicolumn{4}{c}{\text{CODATA 2018}}& 46 & 60 & 1.9 & 0.0019 & 0.00074 \\
\hline
\end{array}
$
    \caption{Corrected Table X. Input data are: one frequency of HD$^+$ (a), one frequency of H$_2^+$ (d), one frequency of D$_2^+$ (f), and H(1S-2S) and H-D(1S-2S).}
    \label{tab:Corrected Table X.}
\end{table}

\eject
\vfill\newpage

\bibliographystyle{elsarticle-num}

\vfill

\end{document}